%% file: acl_latex.tex
\pdfoutput=1
\documentclass[11pt]{article}

\usepackage[final]{acl}
\usepackage{times}
\usepackage{latexsym}
\usepackage[T1]{fontenc}
\usepackage[utf8]{inputenc}
\usepackage{microtype}
\usepackage{inconsolata}
\usepackage{graphicx}
\usepackage{booktabs}
\usepackage{xcolor}
\usepackage{svg}
\usepackage{tabularx}
\usepackage{siunitx}
\usepackage{float}
\usepackage{multirow} 
\usepackage{adjustbox} 
\usepackage{makecell} 
\usepackage{enumitem}
\usepackage[colaction]{multicol}
\usepackage{hyperref}
\usepackage{verbatim}
\usepackage{subfiles}
\usepackage{caption}
\usepackage{subcaption}
\usepackage{rotating}
\usepackage{makecell}
\usepackage{pifont}
\usepackage{url}
\usepackage{hyperref}
\usepackage{amssymb}

\newcommand{\cmark}{\textcolor{green!60!black}{\ding{51}}} 
\newcommand{\xmark}{\textcolor{red!70!black}{\ding{55}}}   

\newcommand{\ourlogo}{\raisebox{-0.42em}{\includegraphics[height=2em]{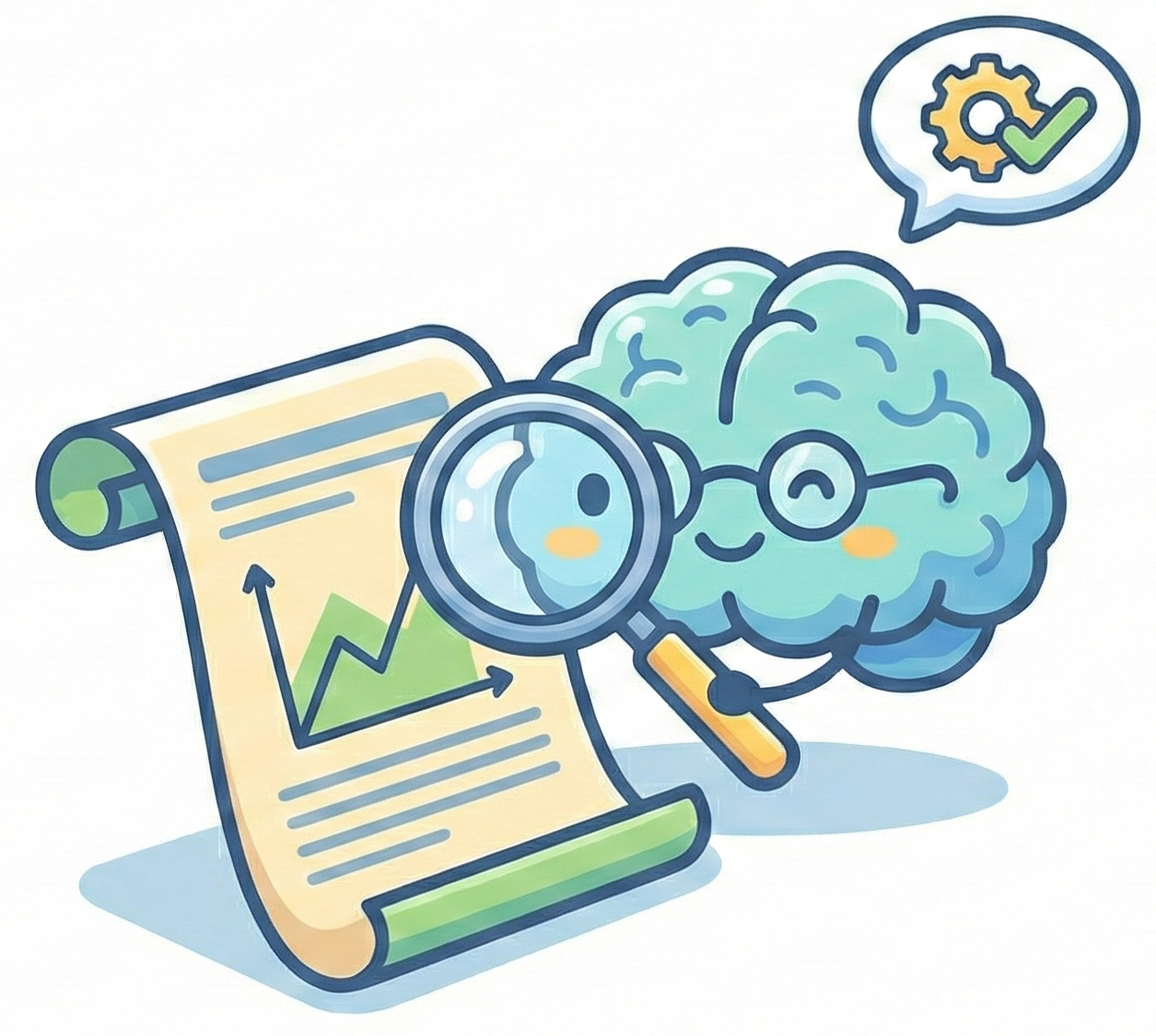}}}
\newcommand{\name}{PaperMind}
\providecommand{\Platform}{\name}

\title{\ourlogo\ \name: Benchmarking Agentic Reasoning and Critique over Scientific Papers in Multimodal LLMs}

\author{{\normalfont Yanjun Zhao\footnotemark[1],} {\normalfont Tianxin Wei\footnotemark[1],} {\normalfont Jiaru Zou,} {\normalfont Xuying Ning,} {\normalfont Yuanchen Bei}\\
{Lingjie Chen}, {Simmi Rana},  {Wendy H. Yang}, {Hanghang Tong}, {Jingrui He}$^{\dag}$\\
\vspace{-0.8em} \\
University of Illinois Urbana-Champaign\\
\texttt{ \{yanjunzh, jingrui\}@illinois.edu}
}
\begin{document}
\maketitle

{
\renewcommand{\thefootnote}{\fnsymbol{footnote}}
\footnotetext[1]{Equal contribution. $^\dag$Corresponding author.}
}

\begin{abstract}

Understanding scientific papers requires more than answering isolated questions or summarizing content. It involves an integrated reasoning process that grounds textual and visual information, interprets experimental evidence, synthesizes information across sources, and critically evaluates scientific claims. However, existing benchmarks typically assess these abilities in isolation, making it difficult to evaluate scientific paper understanding as a unified set of interacting cognitive abilities. In this work, we introduce \name, a benchmark designed to evaluate integrated and agent-oriented scientific reasoning over research papers. \name~is constructed from real scientific papers across seven domains, including agriculture, biology, chemistry, computer science, medicine, physics, and economics. It comprises four complementary task families that collectively operationalize distinct cognitive facets of scientific paper reasoning, including multimodal grounding, experimental interpretation, cross-source evidence reasoning, and critical assessment. By analyzing model behavior across multiple tasks, \name~enables a diagnostic evaluation of integrated scientific reasoning behaviors that are difficult to assess through isolated task evaluations. Extensive experiments on both open-source and closed-source multimodal LLMs reveal consistent performance gaps across tasks, highlighting persistent challenges in integrated scientific reasoning and critique. Our benchmark and dataset are available at \url{https://github.com/Yanjun-Zhao/PaperMind}.
\end{abstract}

\begin{figure*}[t]
    \centering
    \includegraphics[width=\textwidth]{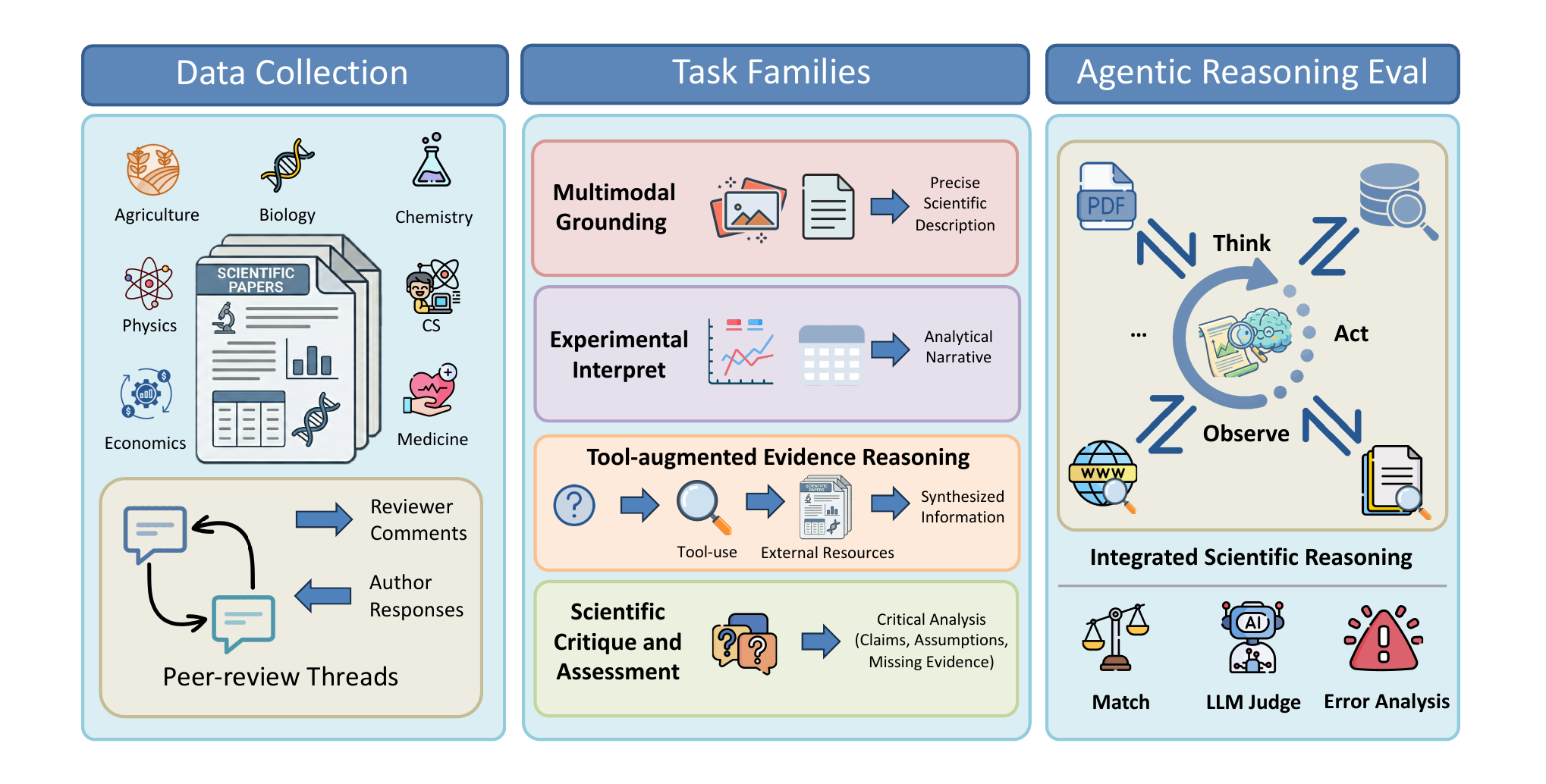}
    \caption{
        Overview of the design and scope of the \name~benchmark.
        }
    \label{fig:benchmark}
\end{figure*}

\section{Introduction}

Understanding scientific literature is a foundational capability for intelligent systems designed to support scientific research. Recent advances in multimodal large language models (LLMs) have enabled promising progress on scientific paper understanding tasks, including document question answering, summarization, and citation-grounded retrieval (e.g., QASPER~\cite{dasigi-etal-2021-dataset}; SciFact~\cite{wadden-etal-2020-fact}; PaperQA~\cite{lala2023paperqa}). In parallel, tool-augmented and agentic LLM systems have demonstrated the ability to perform multi-step reasoning by interleaving language generation with external actions such as search, code execution, and document retrieval (e.g., ReAct~\cite{yao2023react}; Toolformer~\cite{toolformer}; AutoGen~\cite{wu2024autogen}). These advances suggest the potential for multimodal LLM-based agents to assist with scientific research tasks that require structured reasoning over papers, figures, experiments, and citations.

However, existing evaluations of scientific understanding tend to emphasize individual capabilities in relatively isolated and static settings. Current benchmarks~\cite{pramanick2024spiqa, lee2023qasa, li-etal-2024-m3sciqa, Auer2023TheSS, bei2026memgallerybenchmarkingmultimodallongterm, ning2026mcsearchevaluatingenhancingmultimodal} typically focus on specific aspects of scientific understanding: OCR-oriented and layout parsing benchmarks~\cite{Pfitzmann_2022} primarily assess visual perception; scientific QA benchmarks~\cite{sundar2024cpapers, li2024multimodal} evaluate answer extraction from a single paper or a pre-retrieved corpus; summarization benchmarks~\cite{singh-etal-2024-scidqa} target content abstraction; and tool-use benchmarks~\cite{wang2025paperarena} examine general tool-augmented reasoning skills within a single scientific domain.

While existing benchmarks provide useful insights into individual capabilities, they often abstract away the decision-making and coordination behaviors required when models act as agents in scientific analysis. In realistic research workflows, understanding a paper involves grounding figures in context, interpreting experimental results, selectively retrieving and integrating external evidence, and critically assessing claims. These behaviors require models to actively decide what to generate, retrieve, or evaluate, rather than responding to a fixed input in isolation. Consequently, performance on static or single-task benchmarks may not fully reflect model behavior in agentic, end-to-end scientific reasoning settings.

In this work, we introduce a benchmark for scientific paper understanding that evaluates LLMs’ integrated reasoning across multimodal grounding, experimental interpretation, evidence synthesis with tool use, and critical assessment. As illustrated in Figure~\ref{fig:benchmark}, the benchmark comprises four task families. These tasks cover (i) grounding visual content in scientific context and producing precise descriptions, (ii)  interpreting experimental results and generating coherent analytical narratives, (iii) performing agentic reasoning with tool use to retrieve and synthesize evidence from multiple papers or external sources, and (iv) critically examining scientific claims by clarifying ambiguities, identifying weaknesses, missing evidence, or unclear assumptions based on real peer-review scenarios. Rather than treating these abilities in isolation, the benchmark evaluates how models perform across them, probing analysis, synthesis, and judgment behaviors that arise in realistic scientific workflows, and aligning with growing interest in agentic and tool-augmented reasoning systems.

\input{Table/comparison_existed}

Our core contributions are as follows: 
\begin{itemize}[leftmargin=*, itemsep=0pt]
\item \textbf{Benchmark Construction:} We introduce a benchmark for scientific paper understanding that evaluates integrated scientific reasoning across multimodal grounding, experimental interpretation, evidence synthesis with tool use, and critical assessment, reflecting the diverse reasoning behaviors required in realistic scientific workflows.

\item \textbf{Scientific Agentic Task Design:} We introduce agent-oriented tasks that cover both tool-augmented evidence synthesis and rebuttal-style critique. The former evaluates evidence retrieval and integration across sources, while the latter is grounded in high-quality, real peer-review discussions, assessing critical reasoning over substantive scientific claims.

\item \textbf{Systematic Evaluation and Analysis:} We conduct extensive evaluations across both open-source and closed-source multimodal LLMs, providing a systematic comparison and diagnostic analysis that reveals performance gaps and characteristic failure modes in scientific reasoning.

\end{itemize}

\section{Related Works}

\subsection{Scientific Paper QA}

Scientific QA and document understanding benchmarks have rapidly evolved toward multi-document and multi-modal evaluation. M3SciQA~\cite{li-etal-2024-m3sciqa} presents a multi-modal, multi-document scientific QA benchmark over cited paper clusters. SciDQA~\cite{singh-etal-2024-scidqa} offers deep reading comprehension over scientific articles with multi-document reasoning demands. PeerQA~\cite{baumgartner-etal-2025-peerqa} leverages real peer-review questions with answers annotated by authors. SQuAI~\cite{squai} studies multi-agent retrieval-augmented generation with explicit citations over large scientific corpora. DocHop-QA~\cite{park2025dochopqamultihopreasoningmultimodal} emphasizes multi-hop reasoning across multimodal scientific documents. SPIQA~\cite{pramanick2024spiqa} evaluates multimodal QA grounded in figures and tables from research papers. QASA~\cite{lee2023qasa} advances QA over full-text scientific articles with detailed question formulation. SCITAT~\cite{zhang-etal-2025-scitat} provides a benchmark for QA over scientific text and tables with diverse reasoning types. Additional resources such as DocGenome~\cite{xia2024docgenome} support structured document parsing and multi-page QA evaluation, and citation-graph-based benchmark efforts~\cite{hu2025cgragresearchquestionanswering, Auer2023TheSS} explore retrieval over scholarly networks. Despite these advances, existing benchmarks typically focus on isolated aspects of scientific question answering, such as single-document comprehension, factual retrieval, or narrowly defined reasoning skills. In contrast, our benchmark unifies diverse scientific QA paradigms by jointly evaluating comprehensive understanding ability of LLM.

\subsection{Tool-Augmented Agentic Reasoning}

Many recent works~\cite{singh2025agenticreasoningtoolintegration, wei2026agenticreasoninglargelanguage, zhao2025secondorderfinetuningpainllmsa} enhance the tool-use capabilities of LLM agents through SFT and RL training~\cite{zhou2026lookinwardexploreoutward, Zhao_Zhao_Song_He_Zhang_Zhang_Li_2026,ren2025riskporiskbasedpolicyoptimization, dang2025fzoofastzerothorderoptimizer}. Approaches such as ReAct~\cite{yao2023react} and PAL~\cite{PAL} interleave natural-language reasoning with actions or program execution, while Toolformer~\cite{toolformer} enables models to autonomously invoke external APIs. MRKL systems~\cite{karpas2022mrkl} further propose modular architectures that combine LLMs with symbolic solvers and knowledge sources. Building on these ideas, frameworks such as AutoGen~\cite{wu2024autogen}, MetaGPT~\cite{hong2024metagpt}, CAMEL~\cite{camel}, and ChatDev~\cite{qian2024chatdevcommunicativeagentssoftware} investigate role-based collaboration and tool use among LLM agents. More recent evaluations include the ToolHop~\cite{ye-etal-2025-toolhop} benchmark for multi-hop tool invocation and the PaperArena~\cite{wang2025paperarena} for tool-augmented scientific literature reasoning. In comparison, our benchmark provides fine-grained analysis of tool-use failure patterns and evaluates models’ abilities to rebut reviewer-style critiques beyond task-level success metrics.

\section{Benchmark Construction}
This section describes the construction of our benchmark, from paper collection and filtering to task design and ground-truth validation.

\subsection{Paper Collection}
We initially collected 3,000 scientific papers from open-access sources including ArXiv~\footnote{https://arxiv.org}, bioRxiv~\footnote{https://www.biorxiv.org}, and Semantic Scholar~\footnote{https://www.semanticscholar.org}, spanning seven scientific domains: agriculture, biology, chemistry, computer science, medicine, physics, and economics. All papers are released under permissive open licenses. These papers are used to construct the first three tasks. For the fourth task (Critical Assessment), we collect publicly available peer review discussions from OpenReview~\footnote{https://openreview.net/}.

\subsection{Filtering Process}
\begin{figure}[t]
    \centering
    \includegraphics[width=\columnwidth]{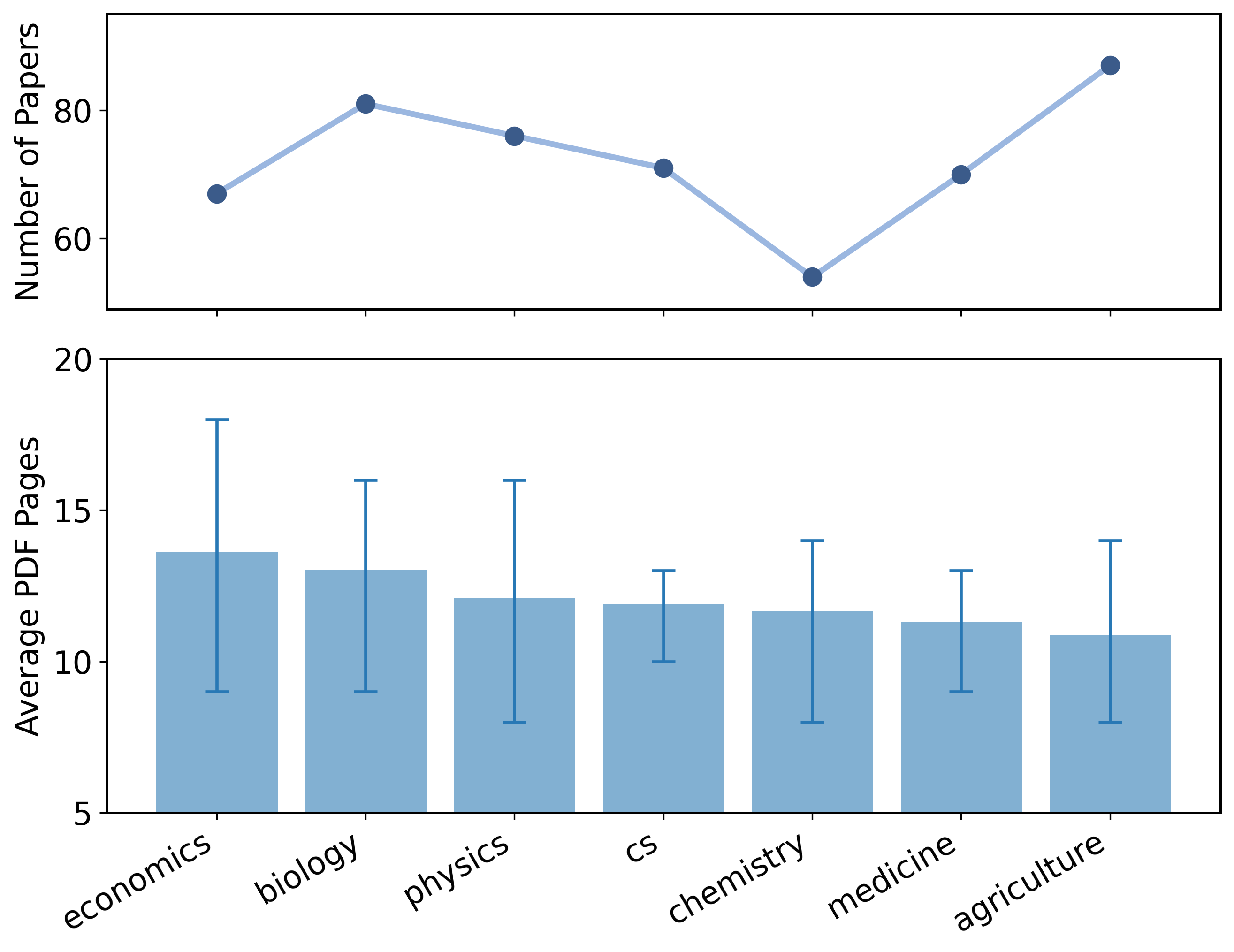}
    \caption{Number of papers (top) and average PDF length in pages (bottom) across different domains.}
    \label{fig:paper_stat}
\end{figure}

We apply a series of filtering and preprocessing steps to ensure data quality. First, we remove papers that are either too short (fewer than five pages) or relatively long (more than twenty pages). We further exclude PDFs with severe formatting issues that hinder reliable parsing. We then employ the paper2pdf tool~\cite{pdf2figure} to convert PDFs into structured representations, including text, figures, and tables, which serve as the basis for question construction. Finally, we post-process the extracted content to remove irrelevant or duplicated text introduced during parsing. Detailed statistics on the number of papers retained in each domain are reported in Figure~\ref{fig:paper_stat}. 

\subsection{QA-pair Construction}
We construct a benchmark for scientific paper understanding consisting of four task families, each derived from real scientific papers and grounded in original paper artifacts, including figures, tables, references, and peer review discussions. Figure~\ref{fig:question_distribution} presents the domain-wise distribution of questions for each task family. We also provide detailed QA examples for each task in Figure~\ref{fig:benchmark_old}. 

\paragraph{Multimodal Ground} In this task, the LLM is provided with a figure extracted from a scientific paper, along with the paper’s introduction as contextual background. LLM is required to generate a concise and accurate caption describing the figure’s content. The original figure caption serves as the groundtruth. This task evaluates models’ ability to ground visual information in scientific context, requiring alignment between visual evidence and domain-specific terminology introduced in the paper.
\input{Table/Q123_results}

\paragraph{Experimental Interpretation} In this task, the LLM is provided with a visual or tabular experimental result from a scientific paper, such as performance curves, comparison plots, or experimental result tables. The introduction section in the paper is also part of the input to provide contextual background, then LLM is required to generate an analytical paragraph suitable for inclusion in the main text of the paper. For groundtruth, we identify corresponding discussion paragraphs by locating text spans that explicitly reference each figure or table in the original paper, and filter out paragraphs that are excessively short or overly long to ensure consistent quality. This task evaluates models’ ability to integrate multimodal information with scientific context, interpret experimental evidence and articulate coherent analytical narratives using appropriate domain-specific language.

 \paragraph{Cross-Source Evidence Reasoning} This task requires the LLM to answer research-oriented questions whose solutions depend on evidence distributed across multiple sources. The required information may originate from the source paper’s textual context, cited reference papers, or external domain knowledge accessed through online resources (e.g., Wikipedia). During inference, the model is provided only with the full source paper in PDF form, and must autonomously locate relevant passages, retrieve information from cited works or external sources via tool use, and synthesize evidence across sources to produce a correct answer. This task evaluates models’ abilities in multi-hop reasoning, cross-source evidence synthesis, and tool-augmented agent behavior under realistic scientific inquiry settings.

To construct this task, we first utilize Gemini 2.5 pro~\cite{comanici2025gemini} with the source paper and identify information-dense statements that describe key methods, results, or claims. These statements often involve domain-specific concepts, entities, or terminology that are underspecified within the local context. Next, we prompt the LLM to retrieve complementary descriptions of these concepts from other parts of the paper, cited reference papers, or external knowledge sources. Based on the relationship between the original statement and the retrieved evidence, we then construct a multi-source question that requires integrating both pieces of information, along with a corresponding answer grounded in the collected materials. All generated questions and answers are followed by careful human revision to ensure correctness and factual consistency. 

\paragraph{Critical Assessment} We collect peer-review discussions from OpenReview and apply strict filtering to retain reviewer questions that elicit substantive author responses and lead to explicit score improvements in later review rounds, ensuring that the retained QA pairs reflect meaningful scientific concerns rather than superficial feedback.

To further improve data quality, we use Gemini 2.5 Pro~\cite{comanici2025gemini} to clean and refine the review threads by removing trivial questions (e.g., minor clarifications or missing citations) and non-informative conversational content (e.g., polite acknowledgments). The resulting QA pairs focus on substantive issues such as experimental limitations, unclear assumptions, methodological choices, and missing evidence.

\section{Experiments}

In this section, we evaluate the overall performance of the LLM on four tasks. We further analyze models’ tool usage behaviors, investigate the failure modes in question–answering failures and provide a case study to illustrate representative model behaviors.

\subsection{Experiment Setup}
\paragraph{Implementation Details.}
For closed-source model, we adopt Gemini 2.5 pro~\cite{comanici2025gemini}, Claude 3.5 Sonnet~\cite{anthropic2024claude}, Claude 3 Haiku~\cite{anthropic2024claude}, GPT-4o mini~\cite{hurst2024gpt} for the evaluation. For open-source model, we adopt Qwen3-VL-4B-Instruct~\cite{yang2025qwen3technicalreport}, Gemma-3.1-4B-Instruct~\cite{gemmateam2025gemma3technicalreport}, Phi-3.5-vision-instruct~\cite{abdin2024phi3technicalreporthighly}. 

We evaluate model performance using both F1 score and an LLM-as-a-Judge metric on a 5-point scale. More results such as the variance of F1 and accuracy derived from LLM-as-a-Judge are provided in the Appendix~\ref{appendix:additional_exp}.

For Cross-Source Evidence Reasoning and Critical Assessment task, which involve tool usage, we follow the experimental setup in prior work~\cite{wang2025paperarena} and enable LLMs to interact with external tools using the \texttt{smolagents}~\cite{smolagents} framework together with the \texttt{ReAct}~\cite{yao2022react} paradigm. We also report the average interaction steps and the average number of tool usage to characterize models’ agentic behaviors during reasoning. Additional details regarding model configurations and evaluation settings can be found in the Appendix~\ref{appendix:eval_detail}.

\subsection{Main Results}

\input{Table/Q4_results}

\input{Table/Q1_ablation}

\begin{figure*}[t]
    \centering
    \includegraphics[width=0.9\textwidth]{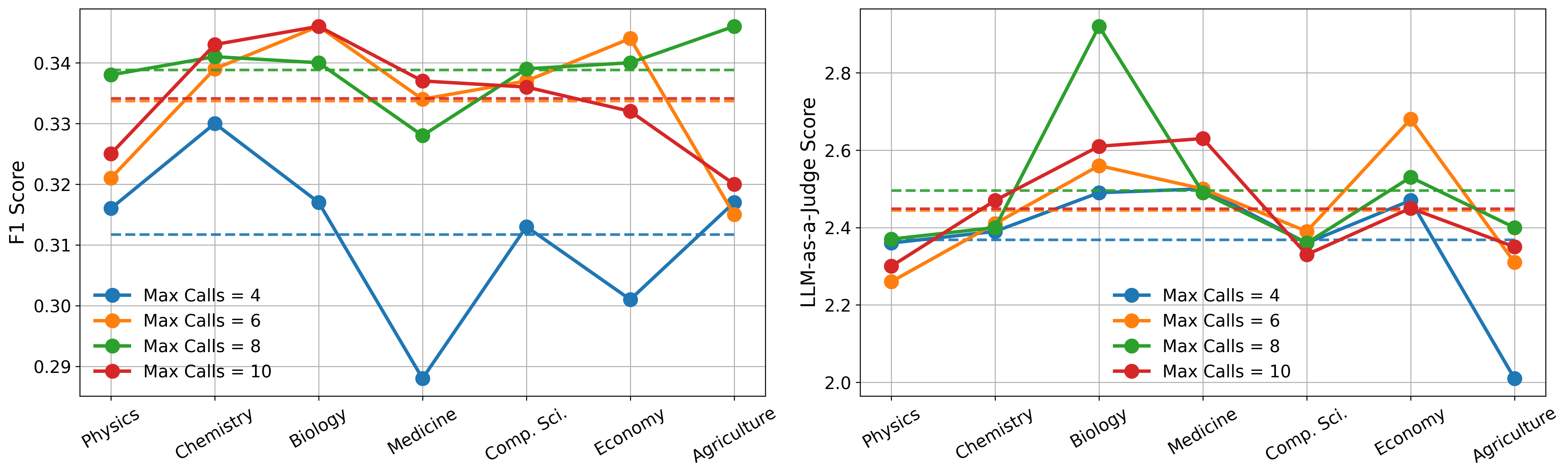}
    \caption{Impact of Maximum Number of Tool Calls on Cross-Source Evidence Reasoning. Dashed lines indicate the average performance across all domains for each setting. 
    }
    \label{fig:tool_number}
\end{figure*}

\begin{figure}[t]
    \centering
    \includegraphics[width=\linewidth]{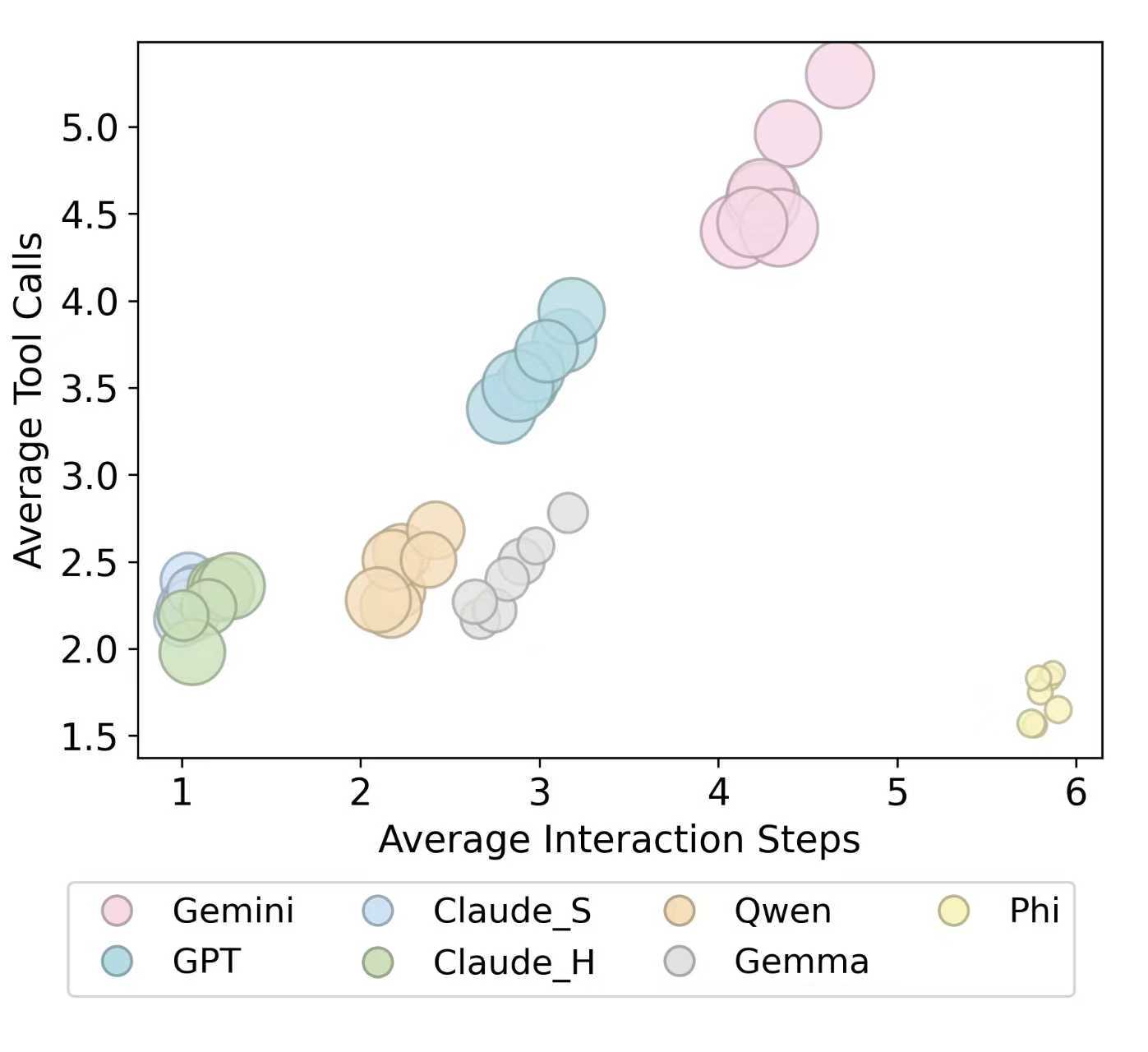}
    \caption{
        Interaction depth versus tool usage of different base LLMs on Cross-Source Evidence Reasoning task across scientific domains.
Each point shows the domain-level average, with colors indicating base LLMs and bubble sizes reflecting the LLM-judge score.
}
    \label{fig:tool_analyze}
\end{figure}

Table~\ref{tab:Q123_results} summarizes the evaluation results on Tasks across different models. For Multimodal Ground, Gemini~2.5~Pro achieves the strongest overall performance, outperforming comparably sized Claude models by 20.4\% in F1 score and 9.6\% in LLM-as-a-Judge ratings. 
For Experimental Interpretation, although Qwen3-VL-4B-Instruct attains higher F1 scores than the Claude 3.5 Sonnet in some cases, Claude consistently achieves better performance under the LLM-as-a-Judge evaluation. Cross-Source Evidence Reasoning task demonstrates more performance gaps observed between closed-source and open-source models. For evaluation results of Critical Assessment in Table~\ref{tab:Q4_results}, Gemini~2.5~Pro achieves the highest LLM-as-a-Judge score, and is also characterized by longer interaction trajectories and more frequent tool usage, indicating a stronger tendency to actively explore evidence and iteratively refine its answers. In contrast, Phi-3.5-vision-instruct performs a relatively large number of interaction steps, but its tool usage frequency remains notably low, causing lower answer quality despite longer reasoning traces.

\input{Table/Q3_input_reference}

\subsection{In-Depth Discussion}
\subsubsection{Effect of Background Introduction on Scientific Tasks}

We investigate the role of background knowledge in scientific figure understanding, with results summarized in Table~\ref{tab:Q1_ablation} for Multimodal Ground. Incorporating introduction-level context leads to consistent improvements in F1 scores for both Gemini 2.5 Pro~\cite{comanici2025gemini} and Qwen3-VL-4B-Instruct~\cite{yang2025qwen3technicalreport}, with gains of approximately 13.7\% and 14.5\%, respectively.
In terms of LLM-as-a-Judge scores, Qwen exhibits a notable improvement of around 8.8\%, whereas Gemini shows only marginal change. This contrast suggests that Gemini possesses stronger inherent generalization ability, enabling it to produce high-quality captions even without explicit background context. We also provide experiment result for Experimental Interpretation in Table~\ref{tab:Q2_ablation} in Appendix~\ref{appendix:additional_exp}.

\subsubsection{Tool Usage Analysis}

\begin{figure}[t]
    \centering
    \includegraphics[width=\linewidth]{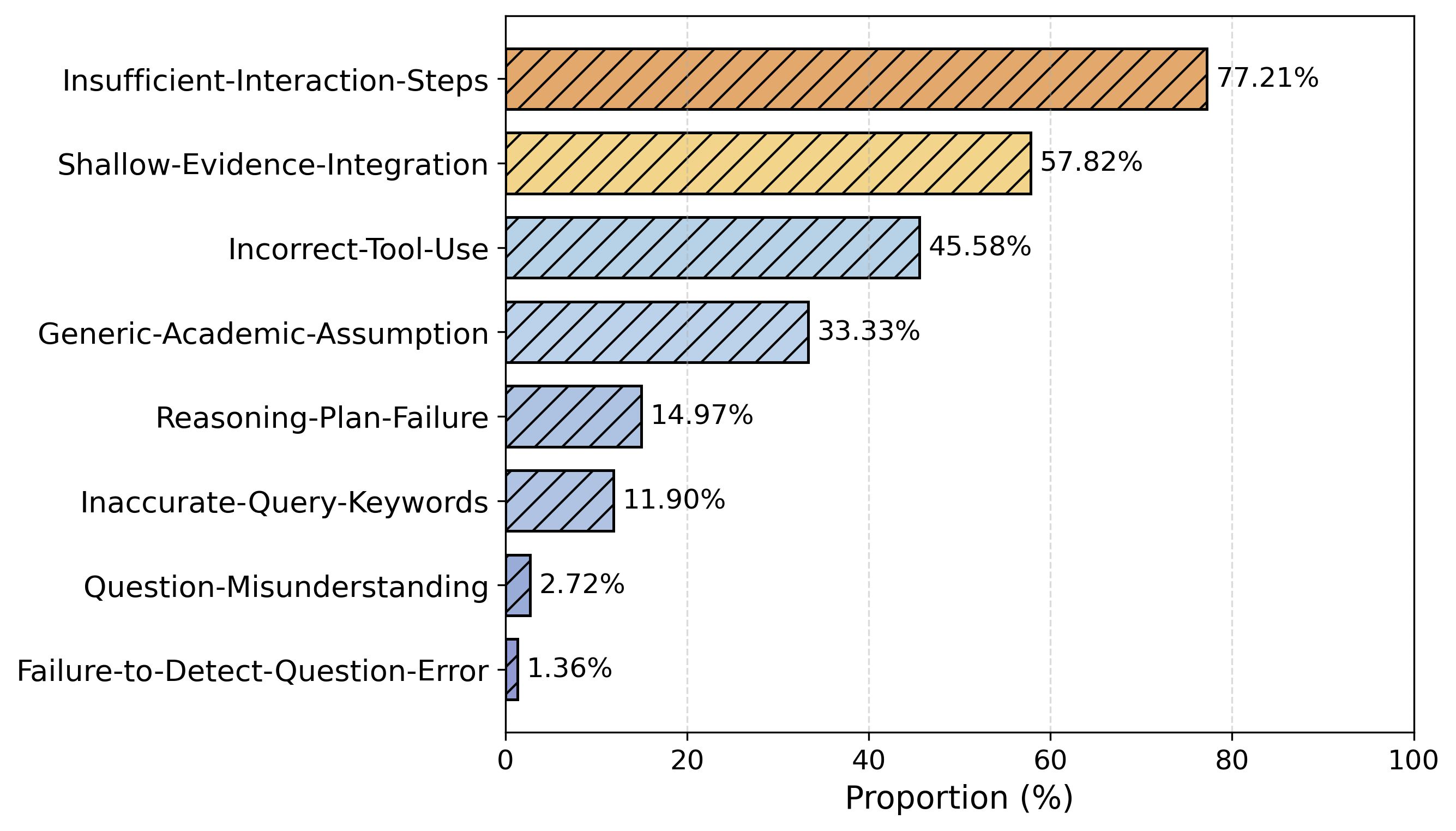}
    \caption{
        Eight-way error taxonomy proportions on Critical Assessment using Qwen3-VL-4B-Instruct (higher is worse), annotated by Gemini-2.5-Pro.}
    \label{fig:error_Q4_qwen}
\end{figure}

\begin{figure}[t]
    \centering
    \includegraphics[width=\linewidth]{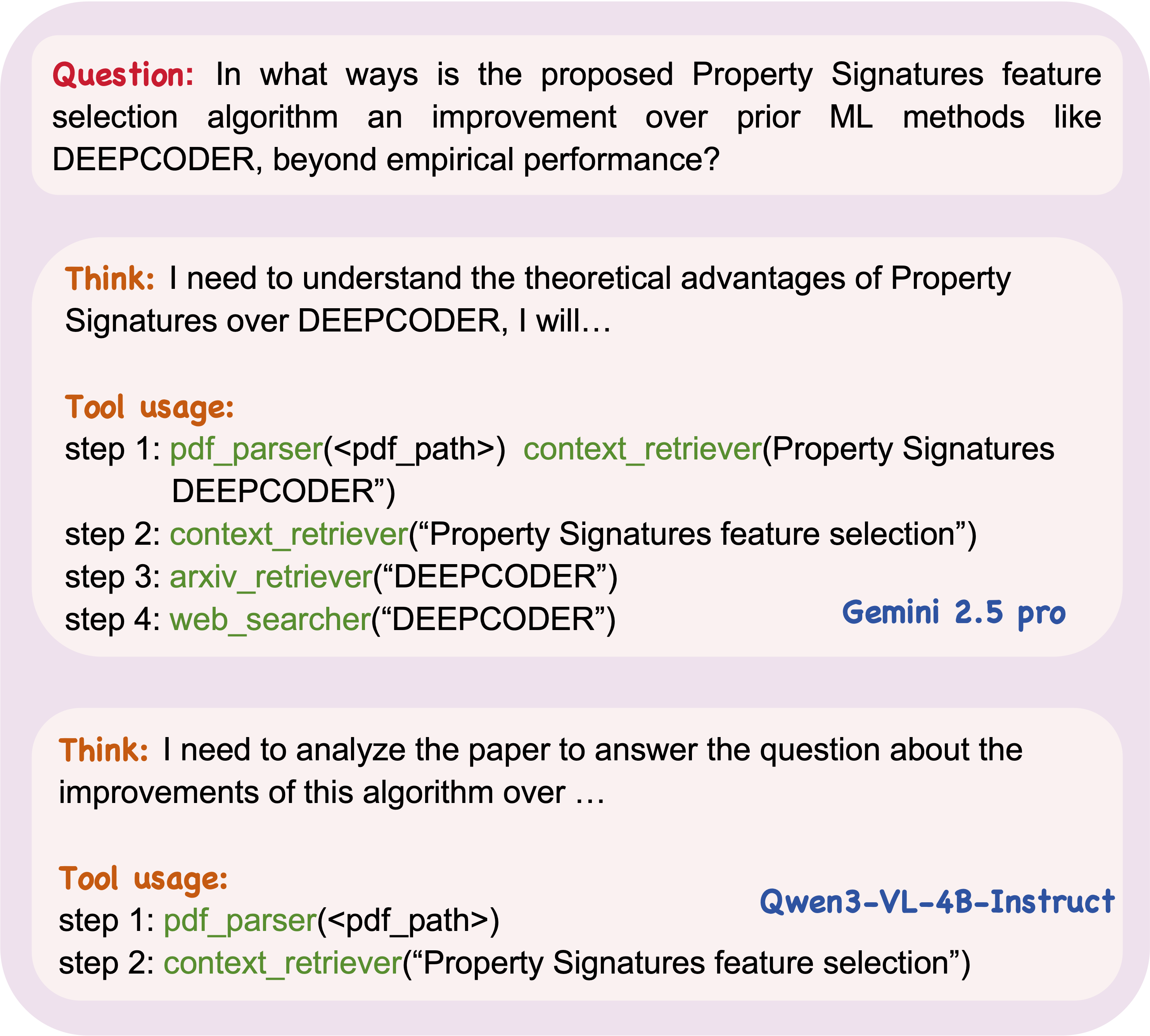}
    \caption{
        A case study illustrating the reasoning process and tool usage of different LLM on Critical Assessment.}
    \label{fig:case_study_Q4}
\end{figure}

We further analyze how different models utilize external tools when solving Cross-Source Evidence Reasoning and Critical Assessment.
Figure~\ref{fig:tool_analyze} illustrates the frequency of tool usage across different models. Overall, we observe that models tend to rely more heavily on general web search, while tool arxiv\_retriever is used less frequently. This suggests that models prefer flexible retrieval strategies that provide broad coverage, rather than strictly targeting specific academic repositories. 

\paragraph{Effect of the Maximum Number of Tool Calls.}
To examine the impact of tool usage depth on performance, we conduct an ablation study by varying the maximum number of allowed tool calls. As shown in Figure~\ref{fig:tool_number}, increasing the tool budget from 4 to 6 leads to the most substantial performance gains across models. Specifically, increasing the tool budget from 6 to 8 yields only marginal gains, and further expanding it to 10 even leads to a slight performance degradation. This trend can be attributed to the increased context length introduced by excessive tool interactions, which may dilute relevant evidence and impose additional burdens on the model’s reasoning and generation processes. More details can be seen in Table~\ref{tab:Q3_different_tool_call} in Appendix~\ref{appendix:additional_exp}.

\paragraph{Validating Multi-Source Reasoning Difficulty.}
We conduct a controlled comparison in which external source information is explicitly provided as part of the LLM input, in contrast to the original formulation where models must independently identify and retrieve relevant external sources through tool use.

As shown in Table~\ref{tab:Q3_input_reference_main}, explicitly providing source information leads to consistent performance improvements across evaluation metrics. The Gemini-2.5-Pro model achieves improvements of 22.6\% and 25.4\% on the F1 score and the LLM-as-a-judge metric, respectively. These gains indicate that access to source identifiers substantially reduces the retrieval burden, enabling models to focus more on evidence integration and answer synthesis. The observed performance gap therefore confirms that the original setting poses a non-trivial multi-source reasoning challenge, as model performance degrades when such reference cues are absent.

\subsubsection{Failure Mode Analysis}
Figure~\ref{fig:error_Q4_qwen} shows the failure mode distribution on Critical Assessment for Qwen3-VL-4B-Instruct. The definition of these error taxonomy and more examples are provided in the Appendix~\ref{appendix:additional_exp}. The most frequent error is Insufficient Interaction Steps (77\%), consistent with the results in Table~\ref{tab:Q4_results}, which indicates that the LLM performs an average of 2.27 interaction steps per question. The second major error source is Shallow Evidence Integration, where retrieved information is not effectively synthesized into a coherent response. In addition, some failures stem from Failure to Detect Question Error, as certain reviewer questions are themselves based on incorrect assumptions. In these cases, Qwen3-VL-4B-Instruct typically does not rebut the flawed premise and instead provides a direct but misaligned answer. 

\subsection{A Case Study}

As shown in the Figure~\ref{fig:case_study_Q4}, the question requires comparing the proposed method with prior approaches in terms of design-level advantages rather than empirical results. Gemini 2.5 pro~\cite{comanici2025gemini} first produces an explicit high-level plan, correctly identifying the need to summarize non-empirical (theoretical or architectural) benefits, while Qwen3-VL-4B-Instruct~\cite{yang2025qwen3technicalreport} directly proceeds to tool invocation without such planning. Both models use basic tools such as PDF parsing and contextual retrieval; however, Gemini further invokes arXiv and web search tools to gather additional evidence, whereas Qwen relies only on document-level information. This example illustrates Gemini’s more deliberate planning and more exhaustive tool usage in complex paper-based QA. 

\section{Conclusion}

We present the PaperMind benchmark for comprehensive scientific paper understanding that evaluates LLM-based systems across four interdependent task families: Multimodal Ground; Experimental Interpretation; Cross-Source Evidence Reasoning and Critical Assessment. We perform large-scale evaluations on both open-source and closed-source LLMs, uncovering substantial performance disparities and recurring failure modes. Compared with existing benchmarks, our proposed benchmark moves beyond isolated retrieval and summarization to assess higher-level reasoning required for scientific research workflows. We believe this work facilitates more systematic evaluation and development of agentic systems for scientific literature understanding.

\section*{Acknowledgement}
This work is supported by Agriculture and Food Research Initiative (AFRI) grant no. 2020-67021-32799/project accession no.1024178 from the USDA National Institute of Food and Agriculture and NSF (2433308). The views and conclusions are those of the authors and should not be interpreted as representing the official policies of the funding agencies or the government.

\section*{Limitations}

Although we rely exclusively on LLM-as-a-Judge for evaluation, the consistency and stability of such automated judgments remain an open challenge, especially across diverse question types. More details can be seen in Appendix. In particular, the alignment between LLM-based judgments and human preferences may vary across different question types and domains, indicating room for improvement in evaluation stability and granularity.

\bibliography{reference}

\clearpage
\appendix

\input{appendix}

\end{document}

%% file: Table/comparison_existed.tex
\begin{table*}[t]
\centering
\renewcommand{\arraystretch}{0.8}
\setlength{\tabcolsep}{4.5pt}
\caption{
Comparison of our proposed benchmark with existing scientific QA benchmarks. 
}
\small
\begin{tabular}{lcccccc}
\toprule
\textbf{Dataset} &
\textbf{\makecell{ Multi \\ domain}} &
\textbf{\makecell{ Whole PDF \\ Input}} &
\textbf{\makecell{ Real-World \\ Reviewer Question}} &
\textbf{\makecell{ Agentic \\Tool Use}} &
\textbf{\makecell{ Integrated \\Reasoning Tasks}} \\
\midrule
MMCR~\cite{tian2025mmcr}           & \cmark & \xmark & \xmark & \xmark & \xmark  \\
ArXivQA~\cite{li-etal-2024-multimodal-arxiv}
 & \cmark & \xmark & \xmark & \xmark &\xmark  \\
PeerQA~\cite{baumgartner2025peerqa}         & \cmark & \xmark & \cmark & \xmark & \xmark \\
$Re^2$~\cite{zhang2025re} & \xmark & \xmark & \cmark & \xmark & \xmark \\
cPAPERS~\cite{sundar2024cpapers}        & \xmark & \xmark & \xmark & \xmark & \xmark   \\
SPIQA~\cite{pramanick2024spiqa}          & \xmark & \xmark & \xmark & \xmark & \xmark  \\
SciVQA~\cite{borisova-etal-2025-scivqa}    & \xmark & \xmark & \xmark & \xmark & \xmark \\
Paperarena~\cite{wang2025paperarena}    & \xmark & \cmark & \xmark & \cmark & \xmark \\
\textbf{Ours}  & \textbf{\cmark} & \textbf{\cmark} & \textbf{\cmark} & \textbf{\cmark} & \textbf{\cmark} \\

\bottomrule
\end{tabular}
\vspace{4pt}
\label{tab:comparison_existed}
\end{table*}

\begin{figure*}[t]
    \centering
    \includegraphics[width=\textwidth]{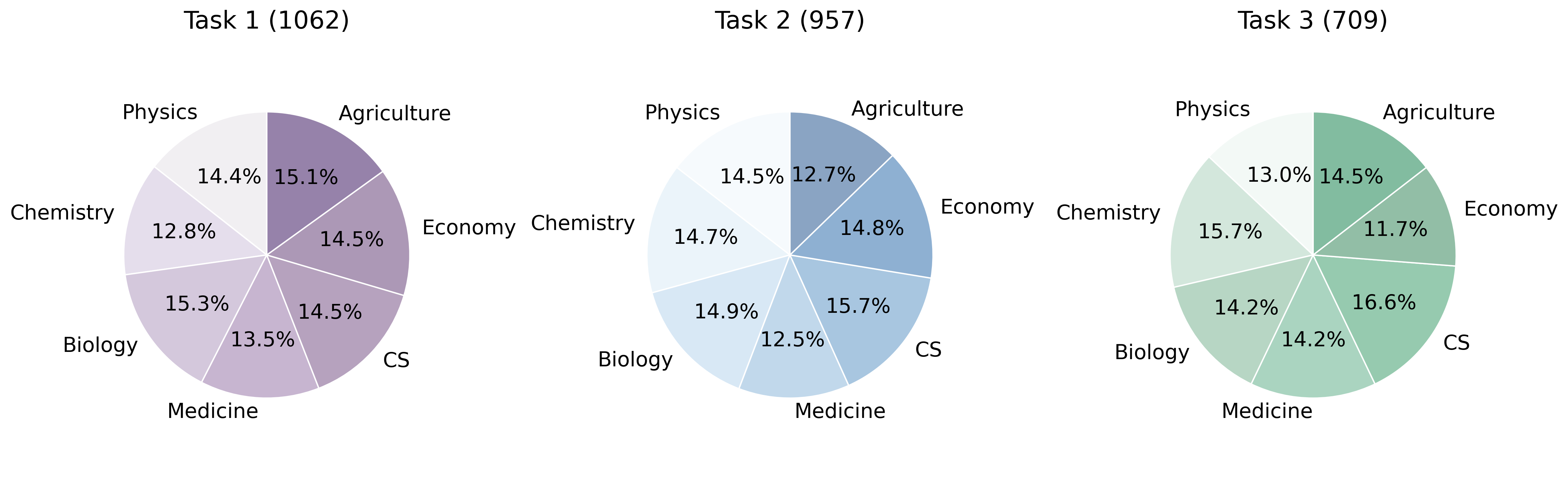}
    \caption{
        Question distribution across seven scientific domains. The benchmark includes four categories of tasks. The pie charts show the distribution of the first three tasks across different scientific domains, annotated with the number of questions and their proportions. The forth Critical Assessment task includes 294 reviewer-author QA pairs from real peer reviews in the computer science domain.
}
    \label{fig:question_distribution}
\end{figure*}

%% file: Table/Q123_results.tex
\begin{table*}[th!]
    \centering
    \setlength{\tabcolsep}{4pt}
    \caption{Evaluation results on Multimodal Ground(MG), Experimental Interpretation(EI) and Cross-Source Evidence Reasoning(ER) task across different models.}
    \label{tab:Q123_results}
    \small
    \resizebox{\textwidth}{!}{
        \begin{tabular}{
            c l | 
            cc cc cc cc cc cc cc | 
            cc
        }
            \toprule
            \multirow{3}{*}{\textbf{Task}}
            & \multirow{3}{*}{\textbf{Base LLM}} 
            & \multicolumn{14}{c|}{\textbf{Scientific Domains}} 
            & \multicolumn{2}{c}{\textbf{Avg. Score}} 
            \\
            \cmidrule(lr){3-16} \cmidrule(lr){17-18}
            &
            & \multicolumn{2}{c}{Physics}
            & \multicolumn{2}{c}{Chemistry}
            & \multicolumn{2}{c}{Biology}
            & \multicolumn{2}{c}{Medical}
            & \multicolumn{2}{c}{Comp. Sci.}
            & \multicolumn{2}{c}{Economy}
            & \multicolumn{2}{c|}{Agriculture}
            & \multirow{2}{*}{F1}
            & \multirow{2}{*}{LLM-J}
            \\
            
            & &F1 & LLM-J
            & F1 & LLM-J
            & F1 & LLM-J
            & F1 & LLM-J
            & F1 & LLM-J
            & F1 & LLM-J
            & F1 & LLM-J
            &
            \\
            \midrule

            \multirow{7}{*}{\rotatebox{90}{\textbf{MG}}}
            &Gemini 2.5 Pro 
                & 0.265 & 2.54
                & 0.223 & 2.38
                & 0.213 & 2.51
                & 0.232 & 2.41
                & 0.220 & 2.33
                & 0.205 & 2.28
                & 0.206 & 2.27 
                & 0.224 & 2.39\\

            &GPT-4o-mini
                & 0.217 & 2.26
                & 0.171 & 2.25
                & 0.196 & 2.35
                & 0.177 & 2.25
                & 0.182 & 2.09
                & 0.156 & 1.97
                & 0.178 & 2.11 
                & 0.183 & 2.18\\

            &Claude 3.5 Sonnet 
                & 0.218  & 2.17
                & 0.174 & 2.22
                & 0.202 & 2.31
                & 0.190 & 2.28
                & 0.183 & 2.02
                & 0.170 & 1.94
                & 0.164 & 2.04
                & 0.186 & 2.14\\
    
            &Claude 3 Haiku
                & 0.210 & 1.81
                & 0.159 & 1.99
                & 0.172 & 1.90
                & 0.168 & 2.03
                & 0.169 & 1.76
                & 0.150 & 1.84
                & 0.153 & 1.79
                & 0.169 & 1.87\\

           &Qwen3-VL-4B-Instruct
                & 0.218 & 2.14
                & 0.180 & 2.08
                & 0.186 & 2.23
                & 0.174 & 2.23
                & 0.178 & 2.00
                & 0.155 & 1.99
                & 0.167 & 2.12
                & 0.181 & 2.11\\

            &gemma-3-4b-it
                & 0.167 & 1.96
                & 0.154 & 1.79
                & 0.167 & 1.88
                & 0.162 & 1.97
                & 0.155 & 1.74
                & 0.206 & 1.92
                & 0.167 & 1.97
                & 0.171 & 1.89\\

            &Phi-3.5-vision-instruct 
                & 0.172 & 1.69
                & 0.163 & 1.78
                & 0.162 & 1.87
                & 0.140 & 1.89
                & 0.170 & 1.75
                & 0.207 & 1.60
                & 0.172 & 1.69 
                & 0.169 & 1.75\\
            \midrule

            \multirow{7}{*}{\rotatebox{90}{\textbf{EI}}}
            & Gemini 2.5 Pro
                & 0.243 & 2.27
                & 0.225 & 2.09
                & 0.191 & 2.00
                & 0.242 & 2.35
                & 0.214 & 2.57
                & 0.224 & 2.00
                & 0.231 & 2.44
                & 0.230 & 2.25\\

            & GPT-4o-mini
                & 0.212 & 1.75
                & 0.197 & 1.80
                & 0.198 & 1.85
                & 0.218 & 1.93
                & 0.202 & 2.12
                & 0.213 & 1.93
                & 0.228 & 2.07 
                & 0.209 & 1.92 \\

            & Claude 3.5 Sonnet
                & 0.215 & 2.04
                & 0.196 & 1.99
                & 0.194 & 2.08
                & 0.202 & 2.14
                & 0.198 & 2.25
                & 0.205 & 2.18
                & 0.214 & 2.20 
                & 0.203 & 2.13 \\

            & Claude 3 Haiku
                & 0.199 & 1.65
                & 0.185 & 1.75
                & 0.172 & 1.70
                & 0.200 & 1.88
                & 0.194 & 1.99
                & 0.195 & 1.92
                & 0.198 & 1.87 
                & 0.192 & 1.82 \\

            & Qwen3-VL-4B-it
                & 0.212 & 1.77
                & 0.197 & 1.76
                & 0.200 & 1.90
                & 0.216 & 1.84
                & 0.206 & 2.03
                & 0.217 & 1.97
                & 0.225 & 2.01
                & 0.210 & 1.90 \\

            & Gemma-3-4b-it
                & 0.205 & 1.55
                & 0.193 & 1.60
                & 0.192 & 1.68
                & 0.224 & 1.68
                & 0.197 & 1.97
                & 0.204 & 1.75
                & 0.216 & 1.86
                & 0.204 & 1.73 \\

            & Phi-3.5-vision-it 
                & 0.192 & 1.45
                & 0.180 & 1.61
                & 0.176 & 1.56
                & 0.194 & 1.63
                & 0.195 & 1.83
                & 0.189 & 1.58
                & 0.193 & 1.59
                & 0.188 & 1.61 \\

            \midrule
            \multirow{7}{*}{\rotatebox{90}{\textbf{ER}}}
            & Gemini 2.5 Pro 
                & 0.317 & 2.77
                & 0.375 & 3.13
                & 0.346 & 3.12
                & 0.320 & 2.84
                & 0.329 & 2.81
                & 0.398 & 3.30
                & 0.358 & 2.94
                & 0.349 & 2.99 \\

            & GPT-4o-mini
                & 0.315 & 2.54
                & 0.283 & 2.64
                & 0.311 & 2.92
                & 0.304 & 2.76
                & 0.290 & 2.51
                & 0.305 & 2.99
                & 0.298 & 2.60
                & 0.301 & 2.71 \\

            & Claude 3.5 Sonnet
                & 0.376 & 2.28
                & 0.386 & 2.34
                & 0.358 & 2.34
                & 0.387 & 2.67
                & 0.382 & 2.21
                & 0.365 & 2.27
                & 0.357 & 1.98
                & 0.373 & 2.30 \\

            & Claude 3 Haiku
                & 0.388 & 2.48
                & 0.380 & 2.40
                & 0.375 & 2.62
                & 0.377 & 2.76
                & 0.366 & 2.31
                & 0.395 & 2.73
                & 0.377 & 2.11
                & 0.380 & 2.49 \\

            & Qwen3-VL-4B-it
                & 0.321 & 2.26
                & 0.339 & 2.41
                & 0.346 & 2.56
                & 0.334 & 2.50
                & 0.337 & 2.39
                & 0.346 & 2.71
                & 0.315 & 2.31
                & 0.337 & 2.45 \\

            & Gemma-3-4b-it
                & 0.191 & 1.73
                & 0.195 & 1.74
                & 0.213 & 1.86
                & 0.218 & 1.96
                & 0.216 & 1.86
                & 0.210 & 1.89
                & 0.180 & 1.64
                & 0.203 & 1.81 \\

            & Phi-3.5-vision-it
                & 0.186 & 1.23
                & 0.194 & 1.24
                & 0.182 & 1.26
                & 0.193 & 1.33
                & 0.193 & 1.36
                & 0.190 & 1.23
                & 0.196 & 1.28
                & 0.191 & 1.28 \\

            \bottomrule
        \end{tabular}
    }
\end{table*}

%% file: Table/Q4_results.tex
\begin{table}[t]
    \centering
    \setlength{\tabcolsep}{5pt}
    \caption{Evaluation results on Critical Assessment task across different models.}
    \label{tab:Q4_results}
    \small
    \begin{tabular}{l | c c c c}
        \toprule
        \multirow{2}{*}{\textbf{Base LLM}} 
        & \multicolumn{4}{c}{\textbf{Comp. Sci.}} \\
        \cmidrule(lr){2-5}
        & \textbf{F1} & \textbf{LLM-J}  & \textbf{Steps} & \textbf{Tools} \\
        \midrule

        \multicolumn{5}{c}{\textit{\Platform{} (Closed-Source Models)}} \\
        \midrule

        Gemini 2.5 Pro 
            & 0.198 & 2.64 &4.58 & 5.37\\

        GPT-4o-mini
            & 0.228 & 2.57 &3.06 &3.75\\

        Claude 3.5 Sonnet
            & 0.228 & 2.43 &1.16 &2.51\\

        Claude 3 Haiku
            & 0.255 & 2.51 &1.15 &2.36\\

        \midrule
        \multicolumn{5}{c}{\textit{\Platform{} (Open-Source Models)}} \\
        \midrule

        Qwen3-VL-4B-Instruct
            & 0.209 & 2.42 &2.27 &2.70\\

        Gemma-3-4b-Instruct
            & 0.146 & 1.95 &2.76 &2.29\\

        Phi-3.5-vision-instruct 
            & 0.134 & 1.26 &5.83 &1.83\\

        \bottomrule
    \end{tabular}
\end{table}

%% file: Table/Q1_ablation.tex
\begin{table*}[th!]
    \centering
    \setlength{\tabcolsep}{4pt}
    \caption{Impact of including the Introduction section of the paper as input on Multimodal Ground task, which serves as background information that helps LLMs better understand the paper.}
    \label{tab:Q1_ablation}
    \small
    \resizebox{\textwidth}{!}{
        \begin{tabular}{
            l | 
            cc cc cc cc cc cc cc | 
            cc
        }
            \toprule
            \multirow{3}{*}{\textbf{Base LLM}} 
            & \multicolumn{14}{c}{\textbf{Scientific Domains}} 
            & \multicolumn{2}{c}{\textbf{Avg. Score}} 
            \\
            \cmidrule(lr){2-15} \cmidrule(lr){16-17}
            & \multicolumn{2}{c}{Physics}
            & \multicolumn{2}{c}{Chemistry}
            & \multicolumn{2}{c}{Biology}
            & \multicolumn{2}{c}{Medical}
            & \multicolumn{2}{c}{Comp. Sci.}
            & \multicolumn{2}{c}{Economy}
            & \multicolumn{2}{c}{Agriculture}
            & \multirow{2}{*}{F1}
            & \multirow{2}{*}{LLM-J}
            \\
            &
            F1 & LLM-J
            & F1 & LLM-J
            & F1 & LLM-J
            & F1 & LLM-J
            & F1 & LLM-J
            & F1 & LLM-J
            & F1 & LLM-J
            &
            \\
            \midrule

            \multicolumn{15}{c}{\textit{\Platform{} (Input with Introduction)}} \\
            \midrule

            Gemini 2.5 Pro 
                & 0.265 & 2.54
                & 0.223 & 2.38
                & 0.213 & 2.51
                & 0.232 & 2.41
                & 0.220 & 2.33
                & 0.205 & 2.28
                & 0.206 & 2.27 
                & 0.224 & 2.39\\
            Qwen3-VL-4B-Instruct
            & 0.218 & 2.14
                & 0.180 & 2.08
                & 0.186 & 2.23
                & 0.174 & 2.23
                & 0.178 & 2.00
                & 0.155 & 1.99
                & 0.167 & 2.12
                & 0.181 & 2.11\\

            \midrule
            \multicolumn{15}{c}{\textit{\Platform{} (Input without Introduction)}} \\
            \midrule
            
            Gemini 2.5 Pro 
                & 0.246 & 2.44
                & 0.223 & 2.44
                & 0.204 & 2.45
                & 0.198 & 2.42
                & 0.199 & 2.14
                & 0.147 & 2.01
                & 0.176 & 2.19 
                & 0.197 & 2.30\\

            Qwen3-VL-4B-Instruct
                & 0.191 & 2.07
                & 0.163 & 2.03
                & 0.166 & 2.15
                & 0.154 & 2.00
                & 0.157 & 1.85
                & 0.130 & 1.69
                & 0.145 & 1.80
                & 0.158 & 1.94\\

            \bottomrule
        \end{tabular}
    }
    \vspace{10pt}

\end{table*}

%% file: Table/Q3_input_reference.tex
\begin{table*}[th!]
    \centering
    \vspace{4pt}
    \setlength{\tabcolsep}{4pt}
    \caption{Effect of explicit external references on Cross-Source Evidence Reasoning task.}
    \label{tab:Q3_input_reference_main}
    \small
    \resizebox{\textwidth}{!}{
        \begin{tabular}{
            l | 
            cc cc cc cc cc cc cc | 
            cc
        }
            \toprule
            \multirow{3}{*}{\textbf{Base LLM}} 
            & \multicolumn{14}{c|}{\textbf{Scientific Domains}} 
            & \multicolumn{2}{c}{\textbf{Avg. Score}} 
            \\
            \cmidrule(lr){2-15} \cmidrule(lr){16-17}
            & \multicolumn{2}{c}{Physics}
            & \multicolumn{2}{c}{Chemistry}
            & \multicolumn{2}{c}{Biology}
            & \multicolumn{2}{c}{Medical}
            & \multicolumn{2}{c}{Comp. Sci.}
            & \multicolumn{2}{c}{Economy}
            & \multicolumn{2}{c|}{Agriculture}
            & \multirow{2}{*}{F1}
            & \multirow{2}{*}{LLM-J}
            \\
            &
            F1 & LLM-J
            & F1 & LLM-J
            & F1 & LLM-J
            & F1 & LLM-J
            & F1 & LLM-J
            & F1 & LLM-J
            & F1 & LLM-J
            &
            \\
            \midrule

            \multicolumn{15}{c}{\textit{\Platform{} (Original Query)}} \\
            \midrule

            Gemini 2.5 Pro 
                & 0.317 & 2.77
                & 0.375 & 3.13
                & 0.346 & 3.12
                & 0.320 & 2.84
                & 0.329 & 2.81
                & 0.398 & 3.30
                & 0.358 & 2.94
                & 0.349 & 2.99 \\
            Qwen3-VL-4B-Instruct
                & 0.321 & 2.26
                & 0.339 & 2.41
                & 0.346 & 2.56
                & 0.334 & 2.50
                & 0.337 & 2.39
                & 0.346 & 2.71
                & 0.315 & 2.31
                & 0.337 & 2.45 \\

            \midrule
            \multicolumn{15}{c}{\textit{\Platform{} (External\_Reference-Augmented Query)}} \\
            \midrule

            Gemini 2.5 Pro 
                & 0.270 & 3.63
                & 0.222 & 4.04
                & 0.216 & 3.79
                & 0.222 & 3.90
                & 0.213 & 3.53
                & 0.168 & 3.99
                & 0.191 & 3.38 
                & 0.428 & 3.75\\
                
            Qwen3-VL-4B-Instruct
                & 0.333 & 2.54
                & 0.342 & 2.78
                & 0.364 & 2.78
                & 0.341 & 2.77
                & 0.359 & 2.69
                & 0.355 & 2.66
                & 0.365 & 2.54
                & 0.351 & 2.68\\

            \bottomrule
        \end{tabular}
    }
\end{table*}

%% file: appendix.tex
\section{Benchmark Construction}
\label{appendix:benchmark}
We provide additional details on paper dataset statistics. We first analyze the page-length distribution of the source papers used in our dataset and report the results in Table~\ref{tab:paper_stat}. This analysis characterizes the diversity and scale of the underlying scientific documents from which our task instances are constructed.

\begin{figure*}[t]
    \centering
    \includegraphics[width=\textwidth]{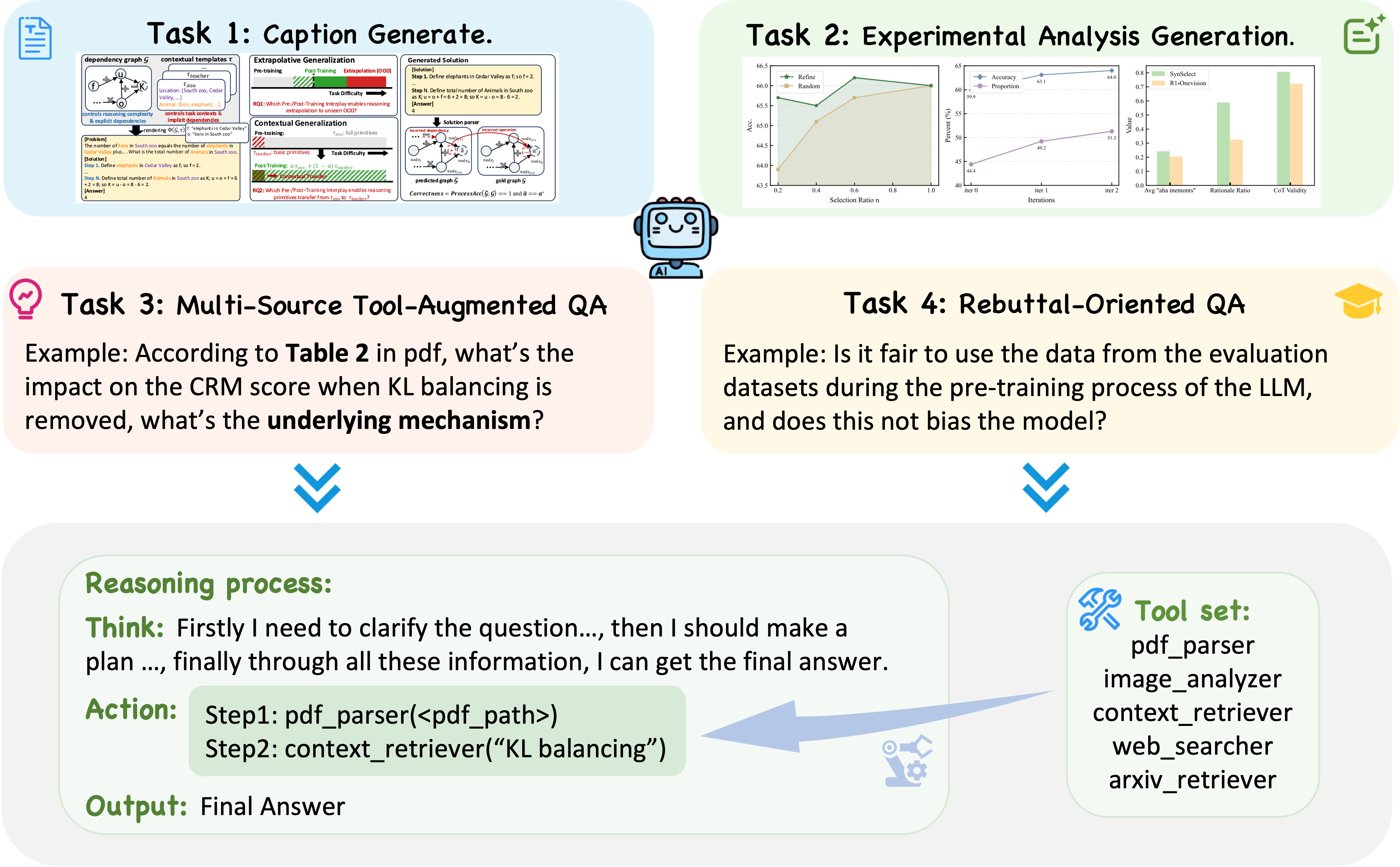}
    \caption{
        Overview of the design and scope of the four tasks.}
    \label{fig:benchmark_old}
\end{figure*}

In addition, we present an illustrating representative task examples together with the corresponding tool libraries required to solve them in Figure~\ref{fig:benchmark_old}.

\section{Evaluation Details}
\label{appendix:eval_detail}

\begin{figure*}[t]
    \centering
    \includegraphics[width=\textwidth]{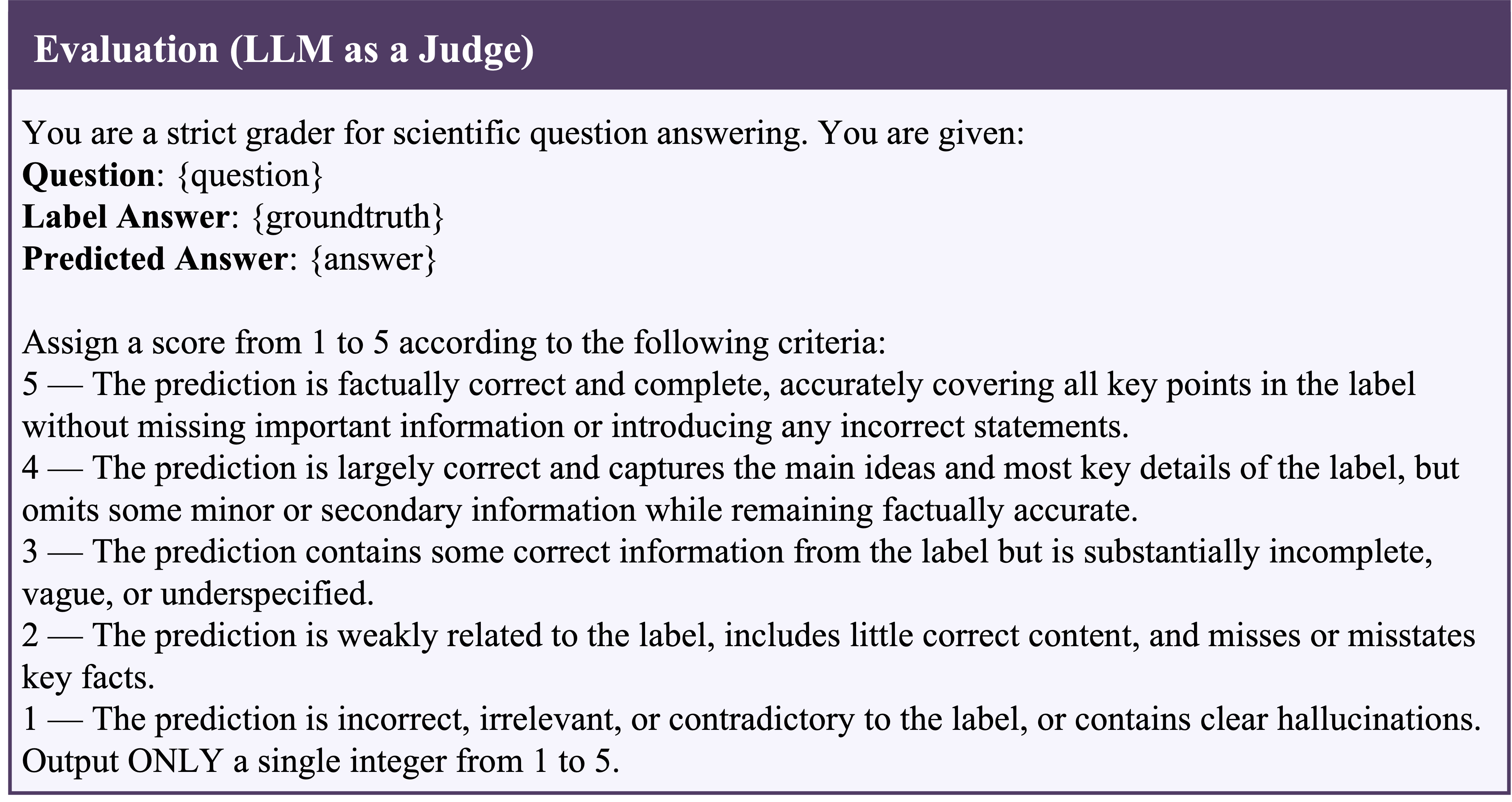}
    \caption{
        Prompt used for LLM-as-a-Judge with GPT-4o, where the model evaluates each prediction on a 5-point scale.}
    \label{fig:prompt_LLM_as_a_judge}
\end{figure*}

All evaluation experiments are conducted on NVIDIA A100 GPU.
Following prior work~\cite{cho2024m3docrag, han2025mdocagentmultimodalmultiagentframework, sun-etal-2025-docagent, li2025rivalreinforcementlearningiterative}, we leverage GPT-4o~\cite{hurst2024gpt} as an LLM-as-a-Judge and F1 to evaluate the consistency between model outputs and reference answers. The LLM-as-a-Judge produces a five-level rating for each prediction based on a rule-based evaluation script that covers different answer types. Detailed prompt can be seen in Figure~\ref{fig:prompt_LLM_as_a_judge}. We report domain-wise averages and variance of the F1 score and LLM-as-a-Judge ratings. Regarding the latter one, we calculate the accuracy by the proportion of predictions receiving a score greater than or equal to 4, which we treat as correct answers.

\section{Additional Experiments}
\label{appendix:additional_exp}


We provide additional experimental results for all four tasks. Specifically, we report the variance of F1 and accuracy calculated from LLM-as-a-judge scores on the Multimodal Ground task (Table~\ref{tab:Q1_LLMjudge_acc}), the Experimental Interpretation task (Table~\ref{tab:Q2_LLMjudge_acc}), the Cross-Source Evidence Reasoning task (Table~\ref{tab:Q3_LLMjudge_acc}), and the Critical Assessment task (Table ~\ref{tab:Q4_acc}).

For deeper analysis, we investigate the impact of input prompts with additional introductions on the Experimental Interpretation task, aiming to examine the effect of background knowledge injection on model performance, shown in Table~\ref{tab:Q2_ablation}.

In the tool usage analysis, we provide detailed statistics on interaction depth and tool usage behaviors of LLM agents on the Cross-Source Evidence Reasoning task in Table~\ref{tab:Q3_tool_stat}, together with an analysis of the impact of the maximum number of tool calls on this task in Table~\ref{tab:Q3_different_tool_call}.

\begin{figure}[t]
    \centering
    \includegraphics[width=\linewidth]{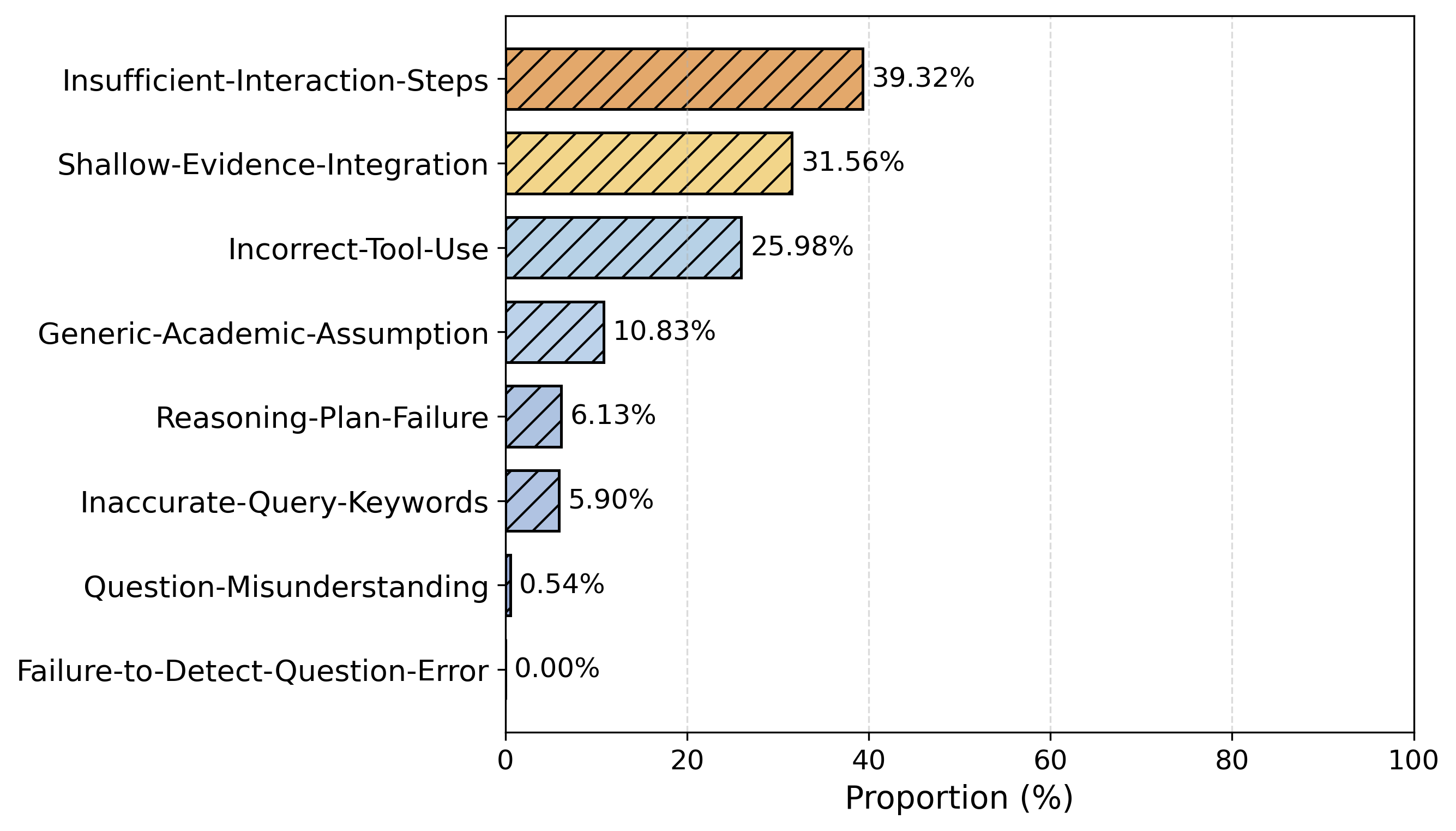}
    \caption{
        Eight-way error taxonomy proportions on Cross-Source Evidence Reasoning using Qwen3-VL-4B-Instruct (higher is worse), annotated by Gemini-2.5-Pro.}
    \label{fig:Q3_error}
\end{figure}

For error analysis, we present the proportions of an eight-way error taxonomy on Cross-Source Evidence Reasoning task using Qwen3-VL-4B-Instruct in Figure~\ref{fig:Q3_error}. We define an eight-category error taxonomy to systematically characterize model failures in complex question answering and tool-augmented reasoning:

(1) Question Misunderstanding refers to cases where the model misundertand the meaning of the question. This error typically arises when the model misinterprets the intent or scope of the question. 

(2) Reasoning-Plan-Failure occurs when the model fails to formulate an appropriate reasoning or action plan before generating intermediate reasoning steps or invoking tools. In such cases, the model output lacks clear sub-goal decomposition, or tool usage proceeds without a coherent strategy, leading to ineffective or misaligned reasoning trajectories.

(3) Incorrect-Tool-Use captures failures where the model selects inappropriate tools or applies them incorrectly for the given information need. This includes choosing an unsuitable tool, misusing tool parameters, or misinterpreting and disregarding the tool outputs during subsequent reasoning steps.

(4) Insufficient-Interaction-Steps describes situations in which the model stops its search process before collecting sufficient evidence to support a well-founded answer. This error is often characterized by an unusually small number of interaction steps, with a final prediction produced despite unresolved uncertainty or incomplete information gathering.

(5) Inaccurate-Query-Keywords refers to cases where the model generates ineffective, overly generic, or misdirected retrieval queries. As a result, the retrieval process repeatedly returns irrelevant, uninformative, or empty results, indicating that the model fails to translate the information need into precise and actionable queries.

(6) Shallow-Evidence-Integration arises when the model successfully retrieves relevant evidence but fails to integrate it into a coherent and well-justified answer. In these cases, the retrieved observations contain the necessary information, yet the final prediction does not adequately combine, explain, or reason over the evidence, falling short of the depth required by the ground-truth label.

(7) Generic-Academic-Assumption characterizes errors where the model relies on generic academic conventions or assumptions rather than concrete evidence from the retrieved materials. This often manifests as reasoning based on common scholarly patterns (e.g., “the paper does not explicitly state…”) instead of grounding conclusions in observed content.

(8) Failure-to-Detect-Question-Error refers to cases in which the model answers a hypothetical or evaluative version of the question, even though the correct response should explicitly state that the paper does not address the issue. This error typically occurs when the question uses normative or evaluative wording, but the model provides speculative analysis rather than correctly reporting the absence of discussion as indicated by the ground-truth label. It often occurs in the peer-review process.

\input{Table/paper_stat}

\begin{figure*}[t]
    \centering
    \includegraphics[width=\textwidth]{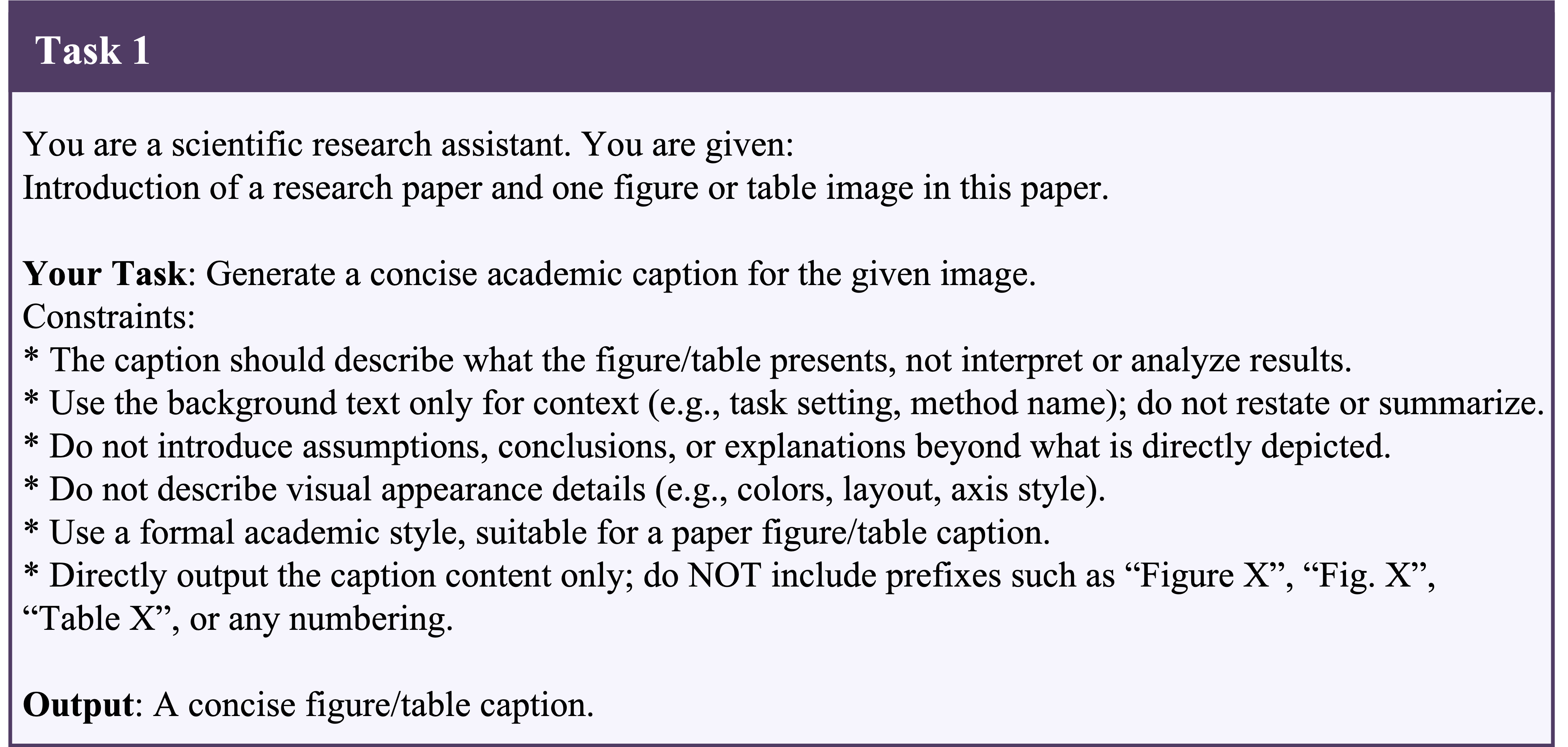}
    \caption{
        Prompt setting with introduction input for Multimodal Ground.}
    \label{fig:task1_prompt}
\end{figure*}

\begin{figure*}[t]
    \centering
    \includegraphics[width=\textwidth]{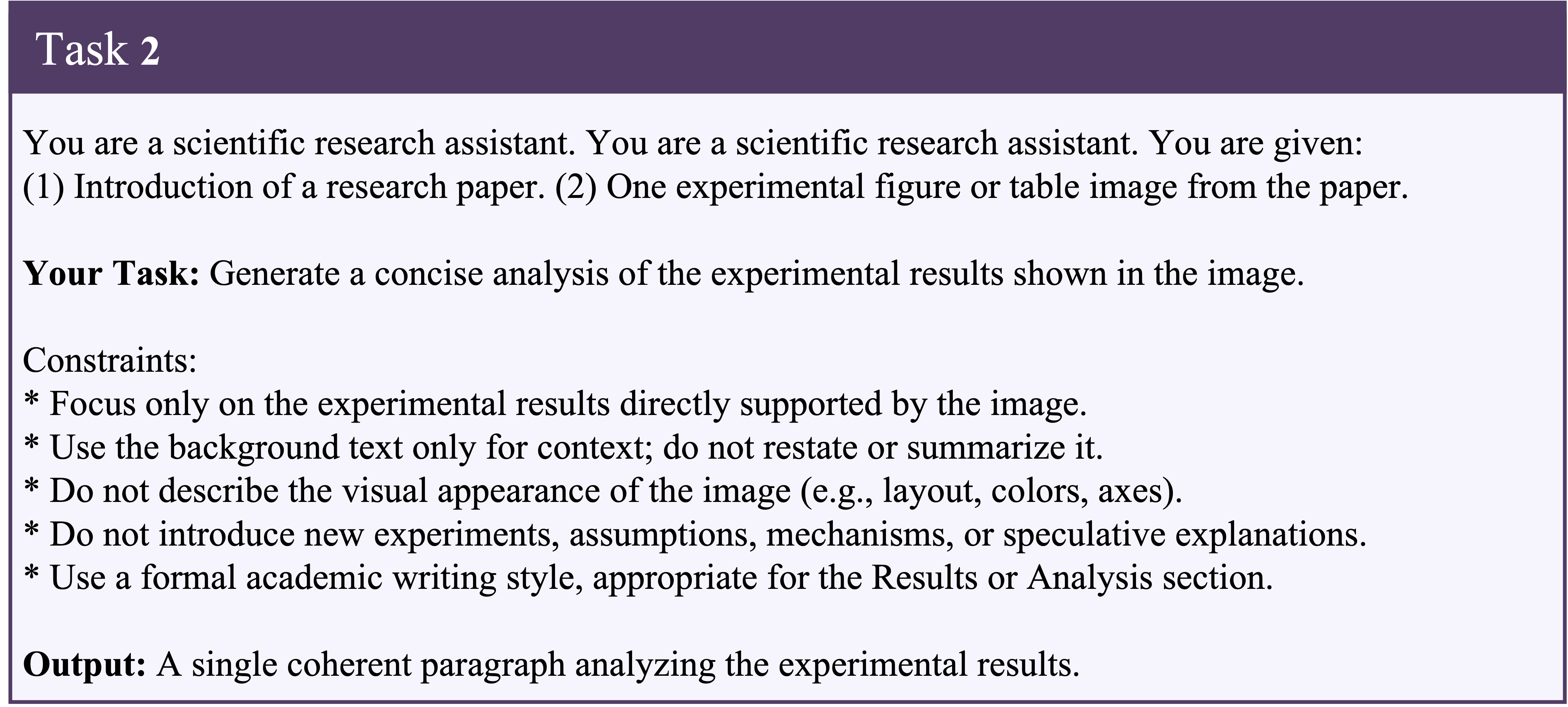}
    \caption{
        Prompt setting with introduction input for Experimental Interpretation.}
    \label{fig:task2_prompt}
\end{figure*}

\begin{figure*}[t]
    \centering
    \includegraphics[width=\textwidth]{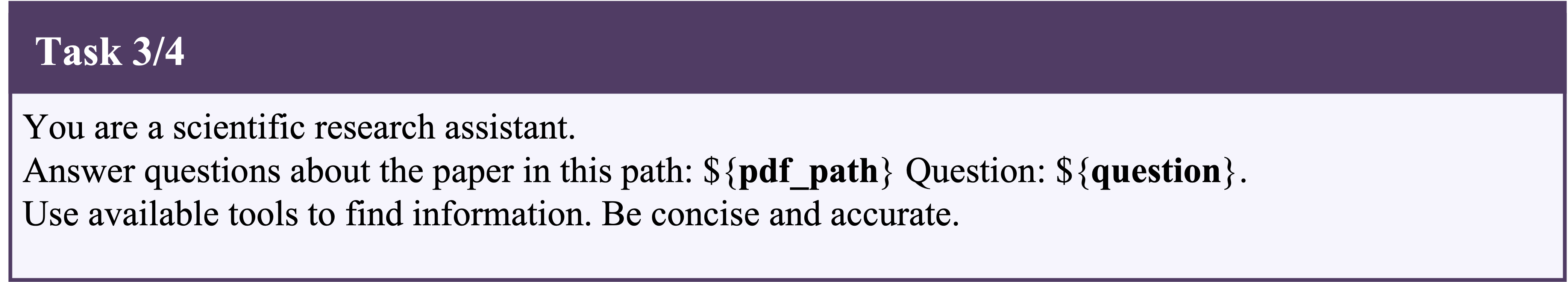}
    \caption{
        Prompt setting for Cross-Source Evidence Reasoning and Critical Assessment. Tool-invocation prompts follow the smolagents framework~\cite{smolagents}.}
    \label{fig:task34_prompt}
\end{figure*}

\input{Table/Q1_LLMjudge_acc}

\input{Table/Q2_LLMjudge_acc}

\input{Table/Q3_LLMjudge_acc}

\input{Table/Q2_ablation}

\input{Table/Q3_tool_stat}

\input{Table/Q3_different_tool_call}

\input{Table/Q3_tool_after_input_reference}

\input{Table/Q4_LLMjudge_acc}

\input{Table/Q4_different_tool_calls}

\section{Ethics Statement}
\label{appendix:ethics}

\subsection{Personal Information Usage}
To ensure data privacy and responsible use, our benchmark is constructed exclusively from publicly available scientific papers sourced from open-access repositories, including arXiv, bioRxiv, Semantic Scholar, and OpenReview. All question–answer pairs are derived from open-source academic content, such as paper text, figures, tables, references, and publicly accessible peer-review discussions.

Throughout the benchmark construction process, all question–answer pairs were carefully reviewed and filtered to remove inaccurate, misleading, or inappropriate content. As the benchmark relies solely on open-access scholarly materials, it does not pose data privacy concerns and is intended solely for research and benchmarking purposes.

\subsection{Human Annotation}
All human involvement in this benchmark was restricted to quality verification. Specifically, manual checks were performed only for Task~3 to verify the correct alignment between multi-source questions and their corresponding ground-truth answers. This process focused solely on factual consistency and evidence matching, without introducing new annotations or subjective judgments. No personal information was collected, and no analysis of annotator behavior was conducted. All verification was carried out by the author team as part of the research workflow, and the benchmark does not involve real user data or human-subject research.

\section{Potential Risks}
\label{appendix:risk}
This benchmark is intended solely for research and evaluation purposes. As it is constructed from scientific papers, potential risks include over-interpreting benchmark performance as a proxy for real-world research competence and propagating biases present in the original academic literature or underlying pretrained models. We mitigate these risks by clearly documenting data sources, task scope, and evaluation limitations, and by positioning the benchmark as a controlled testbed rather than a measure of deployment readiness.

\section{Use of AI Assistants}
In this work, LLMs are employed strictly as supporting tools for benchmark construction and evaluation. During data creation, LLMs are used to assist in generating candidate questions and answers, particularly for multi-source scientific QA, under predefined constraints. These candidates are subsequently manually reviewed and validated by the authors to ensure factual correctness, evidence alignment, and task relevance.

LLMs are also used as an automatic evaluation component through the LLM-as-a-Judge protocol, following established practices in prior research, and in a limited capacity for improving linguistic clarity of the manuscript. All task design, experimental analysis, and scientific conclusions are determined and finalized by the authors, with LLMs serving only as auxiliary tools rather than decision-making agents.

%% file: Table/paper_stat.tex
\begin{table*}[th!]
    \centering
    \setlength{\tabcolsep}{6pt}
    \caption{Dataset statistics across scientific domains, reporting the number of papers and the distribution of PDF page counts. 
    P25 and P75 denote the 25th and 75th percentiles, respectively.}
    \label{tab:paper_stat}
    \small
    \resizebox{\textwidth}{!}{
        \begin{tabular}{
            l |
            c c c c c c c
        }
            \toprule
            \textbf{Statistic} 
            & \multicolumn{7}{c}{\textbf{Scientific Domains}} \\
            \cmidrule(lr){2-8}
            &
            Agriculture
            & Biology
            & Chemistry
            & Comp. Sci.
            & Medicine
            & Physics
            & Economy \\
            \midrule

            \textbf{Paper \#}
            & 87 & 81 & 54 & 71 & 70 & 76 & 67 \\

            \textbf{Avg. Pages}
            & 10.8 & 13.0 & 11.6 & 11.9 & 11.3 & 12.1 & 13.6 \\

            \textbf{P25}
            & 8.0 & 9.0 & 8.0 & 10.0 & 9.0 & 8.0 & 9.0 \\

            \textbf{P75}
            & 14.0 & 16.0 & 14.0 & 13.0 & 13.0 & 16.0 & 18.0 \\

            \bottomrule
        \end{tabular}
    }
\end{table*}

%% file: Table/Q1_LLMjudge_acc.tex
\begin{table*}[th!]
    \centering
    \setlength{\tabcolsep}{4pt}
    \caption{Variance and accuracy of LLM-as-a-judge evaluations on Multimodal Ground task. Scores are assigned on a 5-point scale. Var denotes score variance, and Acc is computed by the percentage treating scores above 4 as success.}
    \label{tab:Q1_LLMjudge_acc}
    \small
    \resizebox{\textwidth}{!}{
        \begin{tabular}{
            l | 
            cc cc cc cc cc cc cc | 
            cc
        }
            \toprule
            \multirow{3}{*}{\textbf{Base LLM}} 
            & \multicolumn{14}{c}{\textbf{Scientific Domains}} 
            & \multicolumn{2}{c}{\textbf{Avg. Score}} 
            \\
            \cmidrule(lr){2-15} \cmidrule(lr){16-17}
            & \multicolumn{2}{c}{Physics}
            & \multicolumn{2}{c}{Chemistry}
            & \multicolumn{2}{c}{Biology}
            & \multicolumn{2}{c}{Medical}
            & \multicolumn{2}{c}{Comp. Sci.}
            & \multicolumn{2}{c}{Economy}
            & \multicolumn{2}{c}{Agriculture}
            & \multirow{2}{*}{Acc}
            & \multirow{2}{*}{Var}
            \\
            
            &Acc & Var
            &Acc & Var
            &Acc & Var
            &Acc & Var
            &Acc & Var
            &Acc & Var
            &Acc & Var
            &
            \\
            \midrule

            \multicolumn{15}{c}{\textit{\Platform{} (Closed-Source Models)}} \\
            \midrule

            Gemini 2.5 Pro 
                & 7.53 & 0.66
                & 10.00 & 0.74
                & 7.64 & 0.61
                & 4.51 & 0.63
                & 5.37 & 0.68
                & 8.11 & 0.72
                & 4.38 & 0.64 
                & 6.79 & 0.67\\

            GPT-4o-mini
                & 4.58 & 0.66
                & 8.09 & 0.78
                & 5.56 & 0.60
                & 4.20 & 0.71
                & 0.65 & 0.53
                & 2.60 & 0.58
                & 3.75 & 0.68
                & 4.20 & 0.65\\

            Claude 3.5 Sonnet 
                & 0.65  & 0.54
                & 6.67 & 0.77
                & 7.64 & 0.71
                & 4.23 & 0.68
                & 1.13 & 0.58
                & 1.95 & 0.53
                & 2.56 & 0.65
                & 3.57 & 0.64\\
    
            Claude 3 Haiku
                & 0.65 & 0.50
                & 2.94 & 0.69
                & 1.23 & 0.58
                & 3.50 & 0.80
                & 0.65 & 0.52
                & 1.30 & 0.60
                & 2.50 & 0.64
                & 1.82 & 0.62\\

            \midrule
            \multicolumn{15}{c}{\textit{\Platform{} (Open-Source Models)}} \\
            \midrule

           Qwen3-VL-4B-Instruct
                & 3.92 & 0.63
                & 2.21 & 0.58
                & 3.09 & 0.62
                & 3.50 & 0.64
                & 2.60 & 0.58
                & 1.30 & 0.55
                & 4.38 & 0.66
                & 3.00 & 0.61\\

            gemma-3-4b-it
                & 1.31 & 0.55
                & 0.74 & 0.52
                & 1.85 & 0.54
                & 1.40 & 0.57
                & 0.65 & 0.51
                & 0.00 & 0.55
                & 1.25 & 0.59
                & 1.03 & 0.54\\

            Phi-3.5-vision-instruct 
                & 1.31 & 0.52
                & 2.94 & 0.62
                & 1.85 & 0.57
                & 2.10 & 0.67
                & 1.30 & 0.51
                & 0.00 & 0.52
                & 0.63 & 0.56
                & 1.45 & 0.57\\

            \bottomrule
        \end{tabular}
    }
\end{table*}

%% file: Table/Q2_LLMjudge_acc.tex
\begin{table*}[th!]
    \centering
    \setlength{\tabcolsep}{4pt}
    \caption{Variance and accuracy of LLM-as-a-judge evaluations on Experimental Interpretation task. Scores are assigned on a 5-point scale. Var denotes score variance, and Acc is computed by the percentage treating scores above 4 as success.}
    \label{tab:Q2_LLMjudge_acc}
    \small
    \resizebox{\textwidth}{!}{
        \begin{tabular}{
            l | 
            cc cc cc cc cc cc cc | 
            cc
        }
            \toprule
            \multirow{3}{*}{\textbf{Base LLM}} 
            & \multicolumn{14}{c}{\textbf{Scientific Domains}} 
            & \multicolumn{2}{c}{\textbf{Avg. Score}} 
            \\
            \cmidrule(lr){2-15} \cmidrule(lr){16-17}
            & \multicolumn{2}{c}{Physics}
            & \multicolumn{2}{c}{Chemistry}
            & \multicolumn{2}{c}{Biology}
            & \multicolumn{2}{c}{Medical}
            & \multicolumn{2}{c}{Comp. Sci.}
            & \multicolumn{2}{c}{Economy}
            & \multicolumn{2}{c}{Agriculture}
            & \multirow{2}{*}{Acc}
            & \multirow{2}{*}{Var}
            \\
            
            &Acc & Var
            &Acc & Var
            &Acc & Var
            &Acc & Var
            &Acc & Var
            &Acc & Var
            &Acc & Var
            &
            \\
            \midrule

            \multicolumn{15}{c}{\textit{\Platform{} (Closed-Source Models)}} \\
            \midrule

            Gemini 2.5 Pro 
                & 7.91 & 0.67
                & 0.00 & 0.5
                & 0.213 & 2.51
                & 15.83 & 1.23
                & 15.33 & 0.70
                & 0.00 & 0.40
                & 12.30 & 0.92 
                & 8.35 & 0.74\\

            GPT-4o-mini
                & 0.72 & 0.34
                & 2.13 & 0.44
                & 3.50 & 0.54
                & 5.00 & 0.70
                & 5.33 & 0.54
                & 2.82 & 0.53
                & 4.91 & 0.69 
                & 3.49 & 0.54\\

            Claude 3.5 Sonnet 
                & 2.16 & 0.33
                & 3.54 & 0.45
                & 2.10 & 0.36
                & 9.16 & 0.62
                & 4.00 & 0.40
                & 7.04 & 0.64
                & 4.92 & 0.61
                & 4.70 & 0.49\\
    
            Claude 3 Haiku
                & 0.00 & 0.27
                & 0.71 & 0.40
                & 0.00 & 0.28
                & 2.50 & 0.49
                & 1.33 & 0.29
                & 4.23 & 0.57
                & 1.64 & 1.79
                & 1.49 & 0.40\\

            \midrule
            \multicolumn{15}{c}{\textit{\Platform{} (Open-Source Models)}} \\
            \midrule

           Qwen3-VL-4B-Instruct
                & 7.91 & 0.39
                & 0.71 & 0.40
                & 2.80 & 0.47
                & 5.00 & 0.75
                & 5.33 & 0.56
                & 3.52 & 0.52
                & 3.28 & 0.55
                & 3.15 & 0.52\\

            gemma-3-4b-it
                & 0.72 & 0.32
                & 1.42 & 0.42
                & 1.40 & 0.33
                & 2.50 & 0.47
                & 1.33 & 0.35
                & 2.11 & 0.50
                & 2.46 & 0.48
                & 1.71 & 0.41\\

            Phi-3.5-vision-instruct 
                & 0.00 & 0.28
                & 0.71 & 0.32
                & 0.70 & 0.34
                & 3.33 & 0.52
                & 0.66 & 0.35
                & 1.41 & 0.46
                & 0.00 & 0.32 
                & 0.97 & 0.37\\

            \bottomrule
        \end{tabular}
    }
\end{table*}

%% file: Table/Q3_LLMjudge_acc.tex
\begin{table*}[th!]
    \centering
    \setlength{\tabcolsep}{4pt}
    \caption{Variance and accuracy of LLM-as-a-judge evaluations on Cross-Source Evidence Reasoning task. Scores are assigned on a 5-point scale. Var denotes score variance, and Acc is computed by the percentage treating scores above 4 as success.}
    \label{tab:Q3_LLMjudge_acc}
    \small
    \resizebox{\textwidth}{!}{
        \begin{tabular}{
            l | 
            cc cc cc cc cc cc cc | 
            cc
        }
            \toprule
            \multirow{3}{*}{\textbf{Base LLM}} 
            & \multicolumn{14}{c}{\textbf{Scientific Domains}} 
            & \multicolumn{2}{c}{\textbf{Avg. Score}} 
            \\
            \cmidrule(lr){2-15} \cmidrule(lr){16-17}
            & \multicolumn{2}{c}{Physics}
            & \multicolumn{2}{c}{Chemistry}
            & \multicolumn{2}{c}{Biology}
            & \multicolumn{2}{c}{Medical}
            & \multicolumn{2}{c}{Comp. Sci.}
            & \multicolumn{2}{c}{Economy}
            & \multicolumn{2}{c}{Agriculture}
            & \multirow{2}{*}{Acc}
            & \multirow{2}{*}{Var}
            \\
            
            &Acc & Var
            &Acc & Var
            &Acc & Var
            &Acc & Var
            &Acc & Var
            &Acc & Var
            &Acc & Var
            &
            \\
            \midrule

            \multicolumn{15}{c}{\textit{\Platform{} (Closed-Source Models)}} \\
            \midrule

            Gemini 2.5 Pro 
                & 38.04 & 2.15
                & 46.85 & 2.09
                & 45.54 & 2.18
                & 38.61 & 2.47
                & 35.59 & 2.11
                & 49.40 & 1.80
                & 39.80 & 2.21 
                & 41.98 & 2.15\\

            GPT-4o-mini
                & 22.83 & 1.34
                & 19.82 & 1.13
                & 26.73 & 1.38
                & 21.78 & 1.01
                & 20.33 & 1.45
                & 30.12 & 1.41
                & 19.42 & 1.15 
                & 23.01 & 1.27\\

            Claude 3.5 Sonnet 
                & 18.48  & 1.46
                & 18.92 & 1.66
                & 17.82 & 1.69
                & 29.70 & 1.49
                & 15.25 & 1.32
                & 18.07 & 1.83
                & 13.59 & 1.48
                & 18.83 & 1.56\\
    
            Claude 3 Haiku
                & 25.00 & 1.75
                & 21.62 & 1.68
                & 26.73 & 1.70
                & 30.69 & 1.71
                & 17.80 & 1.35
                & 32.53 & 1.71
                & 12.62 & 1.28
                & 23.86 & 1.60\\

            \midrule
            \multicolumn{15}{c}{\textit{\Platform{} (Open-Source Models)}} \\
            \midrule

           Qwen3-VL-4B-Instruct
                & 16.30 & 1.32
                & 16.22 & 1.45
                & 17.82 & 1.32
                & 19.80 & 1.42
                & 18.64 & 1.32
                & 26.51 & 1.53
                & 15.53 & 1.19
                & 18.69 & 1.36\\

            gemma-3-4b-it
                & 4.34 & 0.83
                & 5.41 & 0.95
                & 7.92 & 1.05
                & 4.95 & 1.01
                & 6.78 & 0.90
                & 4.81 & 0.84
                & 2.91 & 0.68
                & 5.31 & 0.89\\

            Phi-3.5-vision-instruct 
                & 1.09 & 0.33
                & 2.7 & 0.42
                & 0.00 & 0.33
                & 4.00 & 0.60
                & 2.5 & 0.57
                & 0.0 & 0.30
                & 0.97 & 0.34
                & 1.61 & 0.41\\

            \bottomrule
        \end{tabular}
    }
\end{table*}

%% file: Table/Q2_ablation.tex
\begin{table*}[th!]
    \centering
    \setlength{\tabcolsep}{4pt}
    \caption{Impact of input with Introduction on Experimental Interpretation task.}
    \label{tab:Q2_ablation}
    \small
    \resizebox{\textwidth}{!}{
        \begin{tabular}{
            l | 
            cc cc cc cc cc cc cc | 
            cc
        }
            \toprule
            \multirow{3}{*}{\textbf{Base LLM}} 
            & \multicolumn{14}{c}{\textbf{Scientific Domains}} 
            & \multicolumn{2}{c}{\textbf{Avg. Score}} 
            \\
            \cmidrule(lr){2-15} \cmidrule(lr){16-17}
            & \multicolumn{2}{c}{Physics}
            & \multicolumn{2}{c}{Chemistry}
            & \multicolumn{2}{c}{Biology}
            & \multicolumn{2}{c}{Medical}
            & \multicolumn{2}{c}{Comp. Sci.}
            & \multicolumn{2}{c}{Economy}
            & \multicolumn{2}{c}{Agriculture}
            & \multirow{2}{*}{F1}
            & \multirow{2}{*}{LLM-J}
            \\
            &
            F1 & LLM-J
            & F1 & LLM-J
            & F1 & LLM-J
            & F1 & LLM-J
            & F1 & LLM-J
            & F1 & LLM-J
            & F1 & LLM-J
            &
            \\
            \midrule

            \multicolumn{15}{c}{\textit{\Platform{} (Input with Introduction)}} \\
            \midrule

            Gemini 2.5 Pro
                & 0.243 & 2.27
                & 0.225 & 2.09
                & 0.191 & 2.00
                & 0.242 & 2.35
                & 0.214 & 2.57
                & 0.224 & 2.00
                & 0.231 & 2.44
                & 0.230 & 2.25\\
            Qwen3-VL-4B-Instruct
                & 0.212 & 1.77
                & 0.197 & 1.76
                & 0.200 & 1.90
                & 0.216 & 1.84
                & 0.206 & 2.03
                & 0.217 & 1.97
                & 0.225 & 2.01
                & 0.210 & 1.90 \\

            \midrule
            \multicolumn{15}{c}{\textit{\Platform{} (Input without Introduction)}} \\
            \midrule
            
            Gemini 2.5 Pro
                & 0.206 & 2.07
                & 0.188 & 2.18
                & 0.190 & 2.21
                & 0.203 & 2.32
                & 0.189 & 2.45
                & 0.209 & 2.29
                & 0.231 & 2.42
                & 0.200 & 2.28 \\

            Qwen3-VL-4B-Instruct
                & 0.189 & 1.79
                & 0.175 & 1.88
                & 0.178 & 1.94
                & 0.182 & 1.98
                & 0.169 & 2.15
                & 0.188 & 2.02
                & 0.189 & 2.15
                & 0.181 & 1.99 \\

            \bottomrule
        \end{tabular}
    }
\end{table*}

%% file: Table/Q3_tool_stat.tex
\begin{table*}[th!]
    \centering
    \setlength{\tabcolsep}{4pt}
    \caption{Interaction Depth and Tool Usage Statistics of LLM Agents on Cross-Source Evidence Reasoning task.}
    \label{tab:Q3_tool_stat}
    \small
    \resizebox{\textwidth}{!}{
        \begin{tabular}{
            l | 
            cc cc cc cc cc cc cc | 
            cc
        }
            \toprule
            \multirow{3}{*}{\textbf{Base LLM}} 
            & \multicolumn{14}{c}{\textbf{Scientific Domains}} 
            & \multicolumn{2}{c}{\textbf{Avg. Score}} 
            \\
            \cmidrule(lr){2-15} \cmidrule(lr){16-17}
            & \multicolumn{2}{c}{Physics}
            & \multicolumn{2}{c}{Chemistry}
            & \multicolumn{2}{c}{Biology}
            & \multicolumn{2}{c}{Medical}
            & \multicolumn{2}{c}{Comp. Sci.}
            & \multicolumn{2}{c}{Economy}
            & \multicolumn{2}{c}{Agriculture}
            & \multirow{2}{*}{Steps}
            & \multirow{2}{*}{Tools}
            \\
            
            & Steps & Tools
            & Steps & Tools
            & Steps & Tools
            & Steps & Tools
            & Steps & Tools
            & Steps & Tools
            & Steps & Tools
            &
            \\
            \midrule

            \multicolumn{15}{c}{\textit{\Platform{} (Closed-Source Models)}} \\
            \midrule

            Gemini 2.5 Pro 
                & 4.39 & 4.96
                & 4.11 & 4.40
                & 4.25 & 4.58
                & 4.68 & 5.30
                & 4.24 & 4.62
                & 4.34 & 4.42
                & 4.19 & 4.45
                & 4.31 & 4.68 \\

            GPT-4o-mini
                & 2.93 & 3.52
                & 3.14 & 3.77
                & 2.79 & 3.38
                & 3.18 & 3.94
                & 2.97 & 3.59
                & 2.88 & 3.51
                & 3.04 & 3.71
                & 2.99 & 3.63 \\

            Claude 3.5 Sonnet
                & 1.00 & 2.17
                & 1.04 & 2.39
                & 1.05 & 2.26
                & 1.04 & 2.22
                & 1.08 & 2.33
                & 1.04 & 2.24
                & 1.05 & 2.33
                & 1.04 & 2.28 \\

            Claude 3 Haiku
                & 1.20 & 2.35
                & 1.23 & 2.35
                & 1.23 & 2.34
                & 1.28 & 2.36
                & 1.15 & 2.24
                & 1.06 & 1.98
                & 1.01 & 2.19
                & 1.17 & 2.26 \\

            \midrule
            \multicolumn{15}{c}{\textit{\Platform{} (Open-Source Models)}} \\
            \midrule

            Qwen3-VL-4B-Instruct
                & 2.21 & 2.33
                & 2.23 & 2.55
                & 2.17 & 2.24
                & 2.18 & 2.51
                & 2.42 & 2.68
                & 2.10 & 2.28
                & 2.38 & 2.51
                & 2.24 & 2.44 \\

            Gemma-3-4b-it
                & 2.67 & 2.17
                & 3.16 & 2.78
                & 2.75 & 2.22
                & 2.90 & 2.50
                & 2.82 & 2.40
                & 2.64 & 2.27
                & 2.98 & 2.59
                & 2.85 & 2.42 \\

            Phi-3.5-vision-instruct 
                & 5.85 & 1.83
                & 5.77 & 1.56
                & 5.80 & 1.75
                & 5.90 & 1.65
                & 5.75 & 1.57
                & 5.87 & 1.86
                & 5.79 & 1.83
                & 5.82 & 1.72 \\

            \bottomrule
        \end{tabular}
    }
\end{table*}

%% file: Table/Q3_different_tool_call.tex
\begin{table*}[th!]
    \centering
    \setlength{\tabcolsep}{4pt}
    \caption{Impact of Maximum Number of Tool Calls on Cross-Source Evidence Reasoning task.}
    \label{tab:Q3_different_tool_call}
    \small
    \resizebox{\textwidth}{!}{
        \begin{tabular}{
            l | 
            cc cc cc cc cc cc cc | 
            cc
        }
            \toprule
            \multirow{3}{*}{\textbf{Base LLM}} 
            & \multicolumn{14}{c}{\textbf{Scientific Domains}} 
            & \multicolumn{2}{c}{\textbf{Avg. Score}} 
            \\
            \cmidrule(lr){2-15} \cmidrule(lr){16-17}
            & \multicolumn{2}{c}{Physics}
            & \multicolumn{2}{c}{Chemistry}
            & \multicolumn{2}{c}{Biology}
            & \multicolumn{2}{c}{Medical}
            & \multicolumn{2}{c}{Comp. Sci.}
            & \multicolumn{2}{c}{Economy}
            & \multicolumn{2}{c}{Agriculture}
            & \multirow{2}{*}{F1}
            & \multirow{2}{*}{LLM-J}
            \\
            &
            F1 & LLM-J
            & F1 & LLM-J
            & F1 & LLM-J
            & F1 & LLM-J
            & F1 & LLM-J
            & F1 & LLM-J
            & F1 & LLM-J
            &
            \\
            \midrule

            \multicolumn{15}{c}{\textit{\Platform{} (Max. \#Tool Calls = 4)}} \\
            \midrule

            Qwen3-VL-4B-Instruct
                & 0.316 & 2.36
                & 0.330 & 2.39
                & 0.317 & 2.49
                & 0.288 & 2.50
                & 0.313 & 2.36
                & 0.301 & 2.47
                & 0.317 & 2.01
                & 0.312 & 2.37 \\

            \midrule
            \multicolumn{15}{c}{\textit{\Platform{} (Max. \#Tool Calls = 6)}} \\
            \midrule

            Qwen3-VL-4B-Instruct
                & 0.321 & 2.26
                & 0.339 & 2.41
                & 0.346 & 2.56
                & 0.334 & 2.50
                & 0.337 & 2.39
                & 0.346 & 2.71
                & 0.315 & 2.31
                & 0.337 & 2.45 \\

            \midrule
            \multicolumn{15}{c}{\textit{\Platform{} (Max. \#Tool Calls = 8)}} \\
            \midrule
            
            Qwen3-VL-4B-Instruct
                & 0.338 & 2.37
                & 0.341 & 2.40
                & 0.340 & 2.92
                & 0.328 & 2.49
                & 0.339 & 2.36
                & 0.340 & 2.53
                & 0.346 & 2.40
                & 0.339 & 2.50 \\

            \midrule
            \multicolumn{15}{c}{\textit{\Platform{} (Max. \#Tool Calls = 10)}} \\
            \midrule

            Qwen3-VL-4B-Instruct
                & 0.325 & 2.30
                & 0.343 & 2.47
                & 0.346 & 2.61
                & 0.337 & 2.63
                & 0.336 & 2.33
                & 0.332 & 2.45
                & 0.320 & 2.35
                & 0.334 & 2.45 \\
            \bottomrule
        \end{tabular}
    }
\end{table*}

%% file: Table/Q3_tool_after_input_reference.tex
\begin{table*}[th!]
    \centering
    \setlength{\tabcolsep}{4pt}
    \caption{Effect of explicit external references on Cross-Source Evidence Reasoning task.}
    \label{tab:Q3_input_reference}
    \small
    \resizebox{\textwidth}{!}{
        \begin{tabular}{
            l | 
            cc cc cc cc cc cc cc | 
            cc
        }
            \toprule
            \multirow{3}{*}{\textbf{Base LLM}} 
            & \multicolumn{14}{c}{\textbf{Scientific Domains}} 
            & \multicolumn{2}{c}{\textbf{Avg. Score}} 
            \\
            \cmidrule(lr){2-15} \cmidrule(lr){16-17}
            & \multicolumn{2}{c}{Physics}
            & \multicolumn{2}{c}{Chemistry}
            & \multicolumn{2}{c}{Biology}
            & \multicolumn{2}{c}{Medical}
            & \multicolumn{2}{c}{Comp. Sci.}
            & \multicolumn{2}{c}{Economy}
            & \multicolumn{2}{c}{Agriculture}
            & \multirow{2}{*}{F1}
            & \multirow{2}{*}{LLM-J}
            \\
            &
            F1 & LLM-J
            & F1 & LLM-J
            & F1 & LLM-J
            & F1 & LLM-J
            & F1 & LLM-J
            & F1 & LLM-J
            & F1 & LLM-J
            &
            \\
            \midrule

            \multicolumn{15}{c}{\textit{\Platform{} (Original Query)}} \\
            \midrule

            Gemini 2.5 Pro 
                & 4.39 & 4.96
                & 4.11 & 4.40
                & 4.25 & 4.58
                & 4.68 & 5.30
                & 4.24 & 4.62
                & 4.34 & 4.42
                & 4.19 & 4.45
                & 4.31 & 4.68 \\
            Qwen3-VL-4B-Instruct
                & 2.21 & 2.33
                & 2.23 & 2.55
                & 2.17 & 2.24
                & 2.18 & 2.51
                & 2.42 & 2.68
                & 2.10 & 2.28
                & 2.38 & 2.51
                & 2.24 & 2.44 \\

            \midrule
            \multicolumn{15}{c}{\textit{\Platform{} (External\_Source-Augmented Query)}} \\
            \midrule

            Gemini 2.5 Pro 
                & 3.58 & 4.17
                & 3.74 & 3.86
                & 3.39 & 3.63
                & 3.57 & 4.06
                & 3.50 & 4.18
                & 3.20 & 3.36
                & 3.51 & 4.17
                & 3.50 & 3.92\\
                
            Qwen3-VL-4B-Instruct
                & 2.55 & 2.40
                & 2.56 & 2.73
                & 2.68 & 2.72
                & 2.51 & 2.76
                & 2.53 & 2.61
                & 2.58 & 2.57
                & 2.38 & 2.35
                & 2.54 & 2.59\\

            \bottomrule
        \end{tabular}
    }
\end{table*}

%% file: Table/Q4_LLMjudge_acc.tex
\begin{table}[t]
    \centering
    \setlength{\tabcolsep}{6pt}
    \caption{Variance and accuracy of LLM-as-a-judge evaluations on Critical Assessment task. Scores are assigned on a 5-point scale. Var denotes score variance, and Acc is computed by the percentage treating scores above 4 as success.}
    \label{tab:Q4_acc}
    \small
    \begin{tabular}{l | c c}
        \toprule
        \multirow{2}{*}{\textbf{Base LLM}} 
        & \multicolumn{2}{c}{\textbf{Comp. Sci.}} \\
        \cmidrule(lr){2-3}
        & \textbf{Acc} & \textbf{Var} \\
        \midrule

        \multicolumn{3}{c}{\textit{\Platform{} (Closed-Source Models)}} \\
        \midrule

        Gemini 2.5 Pro 
            & 27.55 & 1.87 \\

        GPT-4o-mini
            & 14.97 & 0.99 \\

        Claude 3.5 Sonnet
            & 11.22 & 0.86 \\

        Claude 3 Haiku
            & 13.26 & 0.94 \\

        \midrule
        \multicolumn{3}{c}{\textit{\Platform{} (Open-Source Models)}} \\
        \midrule

        Qwen3-VL-4B-Instruct
            & 10.2 & 0.82 \\

        Gemma-3-4b-it
            & 3.74 & 0.71 \\

        Phi-3.5-vision-instruct 
            & 0.68 & 0.29 \\

        \bottomrule
    \end{tabular}
\end{table}

%% file: Table/Q4_different_tool_calls.tex
\begin{table}[t]
    \centering
    \setlength{\tabcolsep}{6pt}
    \caption{Impact of Maximum Number of Tool Calls on Critical Assessment task.}
    \label{tab:Q4_different_tool_call}
    \small
    \begin{tabular}{l | c c}
        \toprule
        \multirow{2}{*}{\textbf{Base LLM}} 
        & \multicolumn{2}{c}{\textbf{Comp. Sci.}} \\
        \cmidrule(lr){2-3}
        & \textbf{F1} & \textbf{LLM-J} \\

        \midrule
        \multicolumn{2}{c}{\textit{\Platform{} (Max. \#Tool Calls = 4)}} \\
        \midrule

        Qwen3-VL-4B-Instruct
            & 0.197 & 2.30 \\

        \midrule
        \multicolumn{2}{c}{\textit{\Platform{} (Max. \#Tool Calls = 6)}} \\
        \midrule

        Qwen3-VL-4B-Instruct
            & 0.209 & 2.42 \\

        \midrule
        \multicolumn{2}{c}{\textit{\Platform{} (Max. \#Tool Calls = 8)}} \\
        \midrule

        Qwen3-VL-4B-Instruct
            & 0.212 & 2.44 \\

        \bottomrule
    \end{tabular}
\end{table}